\documentclass[a4paper, 10pt]{article}

\usepackage{geometry}
\usepackage{authblk}
\usepackage{amsthm}
\usepackage[dvipdfmx]{graphicx}
\usepackage{amssymb}
\usepackage{bm}
\usepackage{ascmac}
\usepackage{algorithm,algpseudocode,float}
\usepackage{color}
\usepackage{xspace}

\usepackage[hang,small,bf]{caption}
\usepackage[subrefformat=parens]{subcaption}
\newtheorem{theorem}{Theorem}[section]
\newtheorem{lemma}{Lemma}[section]
\newtheorem{observation}{Observation}[section]
\newtheorem{definition}{Definition}[section]
\newtheorem{corollary}{Corollary}[section]

\newcommand{\TIME}[1]{t_{\fontsize{8pt}{0cm}\selectfont \textit{\texttt{#1}}}}

\title{\textbf{Gathering despite a linear number of weakly Byzantine agents}}

\author[1]{Jion Hirose}
\author[2]{Junya Nakamura}
\author[3]{Fukuhito Ooshita}
\author[1]{Michiko Inoue}

\affil[1]{Nara Institute of Science and Technology}
\affil[2]{Toyohashi University of Technology}
\affil[3]{Fukui University of Technology}

\date{\today}

\begin{document}

\maketitle

\begin{abstract}
We study the gathering problem to make multiple agents initially scattered in arbitrary networks gather at a single node.
There exist $k$ agents with unique identifiers (IDs) in the network, and $f$ of them are weakly Byzantine agents, which behave arbitrarily except for falsifying their identifiers.
The agents behave in synchronous rounds, and each node does not have any memory like a whiteboard.
In the literature, two algorithms for solving the gathering problem have been proposed.
The first algorithm assumes that the number $n$ of nodes is given to agents and achieves the gathering in $O(n^4\cdot |\Lambda_{good}|\cdot X(n))$ rounds, where $|\Lambda_{good}|$ is the length of the largest ID among non-Byzantine agents, and $X(n)$ is the number of rounds required to explore any network composed of $n$ nodes.
The second algorithm assumes that the upper bound $N$ of $n$ is given to agents and at least $4f^2+8f+4$ non-Byzantine agents exist, and achieves the gathering in $O((f+|\Lambda_{good}|)\cdot X(N))$ rounds.
Both the algorithms allow agents to start gathering at different times.
The first algorithm can terminate agents simultaneously, while the second one not.
In this paper, we seek an algorithm that solves the gathering problem efficiently with the intermediate number of non-Byzantine agents since there is a large gap between the numbers of non-Byzantine agents in the previous works.
The resultant gathering algorithm works with at least $8f+8$ non-Byzantine agents when agents start the algorithm at the same time, agents may terminate at different times, and $N$ is given to agents.
To reduce the number of agents, we propose a new technique to simulate a Byzantine consensus algorithm for synchronous message-passing systems on agent systems.
The proposed algorithm achieves the gathering in $O(f\cdot|\Lambda_{good}|\cdot X(N))$ rounds.
This algorithm is faster than the first existing algorithm and requires fewer non-Byzantine agents than the second existing algorithm if $n$ is given to agents, although the guarantees on simultaneous termination and startup delay are not the same.
\end{abstract}

\section{Introduction}
\subsection{Background}
Mobile agents (in short, agents) are software programs that can move autonomously in a distributed system.
A problem to make multiple agents initially scattered in the system meet at a single node is called \emph{gathering}.
This problem is fundamental to various cooperative behavior of agents \cite{Pelc2019book} and allows the agents to exchange information and plan for future tasks efficiently.

Since agents are software programs, they are exposed to bugs, cracking, and other threats.
Thus, as the number of agents increases, it is inevitable that some of those agents become faulty.
Among various faults of agents, Byzantine faults are known to be the most severe because we have no control over the behavior of the faulty agents (called Byzantine agents).
For example, Byzantine agents can stay at the current node, move to a neighbor node, and convey arbitrary information to other agents, deviating from their algorithms.

In this paper, we consider the gathering problem in the presence of Byzantine agents and propose a deterministic synchronous gathering algorithm to solve this problem.

\subsection{Related Works}
\begin{table*}[t]
  \centering
  \footnotesize	
  \caption{A summary of synchronous gathering algorithms in the presence of weakly Byzantine agents assuming that agents have unique IDs.
  Here, input is the information initially given to all agents, $n$ is the number of nodes, $N$ is the upper bound of $n$, $X(n)$ is the number of rounds required to explore any network composed of $n$ nodes, $|\Lambda_{good}|$ is the length of the largest ID among non-Byzantine agents, $|\Lambda_{all}|$ is the length of the largest ID among agents, $k$ is the number of agents, $f$ is the number of Byzantine agents, and $F$ is the upper bound of $f$.}
  \label{tab:RelatedWorks}
  \begin{tabular}{ccccccc}
    \hline
    &Input&\begin{tabular}{c}Condition of \\\#Byzantine agents\end{tabular}&\begin{tabular}{c}Startup\\delay\end{tabular}&\begin{tabular}{c}Simultaneous\\termination\end{tabular}&Time complexity\\
    \hline \hline
    \cite{Dieudonne2014} & $n$ & $f+1\leq k$ & Possible & Possible & $O(n^4\cdot |\Lambda_{good}|\cdot X(n))$\\
    \cite{Dieudonne2014} & $F$ & $2F+2\leq k$ & Possible & Possible & Poly.~of $n$ \& $|\Lambda_{good}|$\\
    \cite{Hirose2021} & $N$ & $4f^2+9f+4\leq k$ & Possible & No guarantee & $O((f+|\Lambda_{good}|)\cdot X(N))$\\
    \cite{Hirose2021} & $N$ & $4f^2+9f+4\leq k$ & Possible & Possible & $O((f+|\Lambda _{all}|)\cdot X(N))$\\
    This study & $N$ & $9f+8\leq k$ & Impossible & No guarantee & $O(f\cdot|\Lambda _{good}|\cdot X(N))$\\
    \hline
  \end{tabular}
\end{table*}

The gathering problem has widely been studied in the literature.
In particular, many of those studies deal with the gathering problem for exactly two agents, which is called the rendezvous problem.
Those studies assume the gathering problem in various environments, which is a combination of the assumptions (e.g., agent synchronization, anonymity, presence/absence of memory on a node (called whiteboard), presence/absence of randomization, topology).
Then, those studies have been clarified the solvability such as the gathering problem and, if solvable, they have been analyzed its cost (e.g., time, the number of moves, memory space, etc.).
Pelc \cite{Pelc2019book} has extensively surveyed deterministic rendezvous problems under the various assumptions.
Also, Alpern et al.\cite{Alpern2006} have described an extensive survey of randomized rendezvous problems under the various assumptions.
In the rest of this section, we describe the existing results for the deterministic gathering in the network, on which we focus in this paper. 

If agents are anonymous (i.e., they do not have IDs) and no whiteboard exists (i.e., agents cannot leave any information on nodes), they cannot achieve the rendezvous for some graphs and initial arrangements because they cannot break the symmetry.
Therefore, several studies \cite{Dessmark2006,Kowalski2008,Ta-Shma2014} break the symmetry by attaching unique IDs to agents, and have clarified the time complexity of the rendezvous algorithm for arbitrary graphs under the assumption that agents behave synchronously in the network.
Dessmark et al.~\cite{Dessmark2006} have provided the rendezvous algorithm in polynomial time of $n$, $|\lambda|$, and $\tau$, where $n$ is the number of nodes, $|\lambda|$ is the length of the smallest ID, and $\tau$ is the delay between the starting time of agents.
Kowalski et al.~\cite{Kowalski2008} and Ta-Shma et al.~\cite{Ta-Shma2014} have proposed the algorithms whose time complexity are independent of $\tau$.
Also, Miller et al.~\cite{Miller2016} have investigated the trade-offs between cost and time required to solve the rendezvous problem.
In contrast, some studies \cite{Fraigniaud2013,Czyzowicz2012how} aim to optimize the memory space, and have investigated the time or the number of moves in the case where agents are anonymous.
In this case, they have provided algorithms for solvable graphs and arrangements.
Fraigniaud et al.~\cite{Fraigniaud2013,Fraigniaud2008} and Czyzowicz et al.~\cite{Czyzowicz2012how} have proposed algorithms for trees and an algorithm for arbitrary graphs, respectively.
Furthermore, Dieudonn\'{e} et al.~\cite{Dieudonne2016} have provided the gathering algorithm in the case where agents are anonymous.

Several studies \cite{Czyzowicz2012how,Bampas2010,Marco2006,Dieudonne2015,Guilbault2013} have considered the rendezvous problem in asynchronous environments (i.e., different agents move at different constant speeds or move asynchronously).
In the latter case, the adversary determines the speed of every agent at each time.
Also, agents cannot achieve the rendezvous possibly, and thus this case allows agents to meet inside an edge.

Recently, some studies \cite{Dieudonne2014,Hirose2021,Bouchard2016,Bouchard2022,Miller2020,Tsuchida2018,Tsuchida2020} have considered the gathering problem in the presence of Byzantine agents assuming that agents have unique IDs, which this study also address.
These studies consider two types of Byzantine agents, weakly and strongly ones.
While weakly Byzantine agents can behave arbitrarily except for falsifying their own IDs, strongly Byzantine agents can behave arbitrarily including falsifying their own IDs.
Table \ref{tab:RelatedWorks} summarizes this study and the related studies in the presence of weakly Byzantine agents.
Note that the assumption of startup delay in Table \ref{tab:RelatedWorks} means agents may start an algorithm at different times but agents can wake up sleeping agents at the visited node.

Dieudonn{\'{e}} et al.~\cite{Dieudonne2014} first introduced the gathering problem in synchronous environments in the presence of weakly Byzantine agents.
They have provided two gathering algorithms under the assumption that $k$ agents exist in an arbitrary network composed of $n$ nodes and at most $F$ of them are Byzantine agents.
The first algorithm solves the gathering problem in $O(n^4\cdot |\Lambda_{good}|\cdot X(n))$ rounds if $k\geq f+1$ holds (i.e., at least one non-Byzantine agent exist) and $n$ is given to agents, where $f$ is the number of Byzantine agents, $|\Lambda_{good}|$ is the length of the largest ID among non-Byzantine agents, and $X(n)$ is the number of rounds required to explore any network composed of $n$ nodes.
The second algorithm achieve the gathering in polynomial time of $|\Lambda_{good}|$ and $X(n)$ if $k\geq 2F+2$ holds (i.e., at least $F+2$ non-Byzantine agent exist) and $F$ is given to agents.
The numbers of non-Byzantine agents used in these algorithms match the lower bounds of the number of non-Byzantine agents required to solve the gathering problem under the assumptions.
Hirose et al.~\cite{Hirose2021} provided the two gathering algorithm with lower time complexity by assuming $\Omega(f^2)$ non-Byzantine agents.
If the upper bound $N$ of $n$ is given to agents and $k\geq 4f^2+9f+4$ holds (i.e., at least $4f^2+8f+4$ non-Byzantine agents exist), the first algorithm achieves the gathering with non-simultaneous termination in $O((f+|\Lambda_{good}|)\cdot X(N))$ rounds, and the second one achieves the gathering with simultaneous termination in $O((f+|\Lambda_{all}|)\cdot X(N))$ rounds, where $|\Lambda_{all}|$ is the length of the largest ID among agents.
Tsuchida et al.~\cite{Tsuchida2018} reduced the time complexity using authenticated whiteboards (i.e., each agent has a dedicated area for each node and can leave the information on its area using its ID).
Their algorithm assumes that $F$ is given to agents and $F<k$ holds and achieves the gathering in $O(Fm)$ rounds, where $m$ is the number of edges.
To efficiently achieve the gathering, the authors proposed a technique for agents to simulate a consensus algorithm for Byzantine message-passing systems.
However, this technique requires each node to have an authenticated whiteboard.
Tsuchida et al.~\cite{Tsuchida2020} have proved that agents achieve the gathering in asynchronous environments by assuming that authenticated whiteboards exist.

Dieudonn{\'{e}} et al.~\cite{Dieudonne2014} also introduced the gathering problem in synchronous environments in the presence of strongly Byzantine agents for the first time and have provided two gathering algorithms under the different assumptions.
The first algorithm solve the gathering problem in exponential of $|\Lambda_{good}|$ and $X(n)$ if $k\geq 3F+1$ holds (i.e., at least $2F+1$ non-Byzantine agents exist) and $n$ and $F$ are given to agents.
The second algorithm achieve the gathering in exponential of $|\Lambda_{good}|$ and $X(n)$ if $k\geq 5F+2$ holds (i.e., at least $4F+2$ non-Byzantine agents exist) and $F$ is given to agents.
On the other hand, the lower bounds on the number of non-Byzantine agents required to solve the gathering problems under these assumptions are $F+1$ and $F+2$, respectively.
Bouchard et al.~\cite{Bouchard2016} have provided two algorithms using the number of non-Byzantine agents that match the lower bounds on the gathering problems under these assumptions.
Bouchard et al.~\cite{Bouchard2022} reduced the time complexity to polynomial time complexity by assuming that $\Omega(f^2)$ non-Byzantine agents exist.
Their algorithm assume that $\lceil \log\log n \rceil$ is given to agents and $k\geq 5f^2+7f+2$ holds (i.e., at least $5f^2+6f+2$ non-Byzantine agents exist), and achieves the gathering in polynomial time of $n$ and $|\lambda_{good}|$, where $|\lambda_{good}|$ is the length of the smallest ID among non-Byzantine agents, and $f$ is the number of Byzantine agents.
Miller et al.~\cite{Miller2020} have proposed the gathering algorithm in small time complexity by additional assumption.
They assume that $k\geq 2f+1$ holds (i.e., $f+1$ non-Byzantine agents exist) and an agent can get the subgraph induced by nodes within distance $D_r$ from its current node and the state of agents in the subgraph, where $D_r$ is the radius of the graph.
Their algorithm achieves the gathering in $O(kn^2)$ rounds.

\subsection{Contribution}
Our goal is to provide an efficient algorithm that achieves the gathering with non-simultaneous termination in synchronous environments where $\Omega(k)$ weakly Byzantine agents exist.
Dieudonn\'{e} et al.~\cite{Dieudonne2014} proposed the first algorithm that achieves the gathering in weakly Byzantine environments.
The algorithm achieves the gathering with simultaneous termination in $O(n^4\cdot |\Lambda_{good}|\cdot X(n))$ rounds under the assumption that the number $n$ of nodes is given to agents and at least $f+1$ agents exist in the network, where $f$ is the number of Byzantine agents, $|\Lambda_{good}|$ is the length of the largest ID among non-Byzantine agents, and $X(n)$ is the number of rounds required to explore any network composed of $n$ nodes.
Hirose et al.~\cite{Hirose2021} proposed an algorithm with lower time complexity by assuming that $\Omega(f^2)$ non-Byzantine agents exist for $f$ Byzantine agents in the network.
The algorithm achieves the gathering with non-simultaneous termination in $O((f+|\Lambda_{good}|)\cdot X(N))$ rounds if the upper bound $N$ of $n$ is given to agents and at least $4f^2+9f+4$ agents exist in the network.
In summary, the former algorithm requires a small number of non-Byzantine agents but has high time complexity, while the latter algorithm requires a large number of non-Byzantine agents but has low time complexity.
In particular, if agents need to achieve the gathering when the number of non-Byzantine agents is $\Omega(f)$, they must only choose the former algorithm with high time complexity.

In this paper, we propose a deterministic gathering algorithm with low time complexity in the existence of $\Omega(f)$ non-Byzantine agents.
Since there is a large gap between the assumptions of the number of agents in the above works, it is reasonable to consider an efficient gathering algorithm under the intermediate assumption.
The proposed algorithm achieves the gathering with non-simultaneous termination in $O(f\cdot |\Lambda_{good}|\cdot X(N))$ rounds under the assumption that $N$ is given to agents, at least $9f+8$ agents exist in the network, and agents start the algorithm at the same time.
This algorithm is faster than that of Dieudonn\'{e} et al.~and requires fewer non-Byzantine agents than that of Hirose et al.~if $n$ is given to agents, although the guarantees on simultaneous termination and startup delay are not the same.
To solve the gathering under these assumptions, we propose a new technique to simulate a consensus algorithm \cite{Khanchandani2021} for synchronous Byzantine message-passing systems on agent systems, in which one agent imitates one process.
It is known that Byzantine consensus is solvable on a synchronous distributed system with at least $3b+1$ processes where $b$ is the number of Byzantine processes \cite{Pease1980,Lamport1982}.
However, it is difficult for all agents to simulate synchronous rounds of Byzantine message-passing systems and start the consensus algorithm at the same time.
Instead, we construct a group of at least $3f+1$ agents that realize the above behavior.
This technique can be used not only for Byzantine gatherings but also for other problems.

\section{Agent Model and Problem}
\subsection{Model}
\paragraph{Agent System}
Agent system is modeled by a connected undirected graph $G=(V,E)$, where $V$ is a set of $n$ nodes and $E$ is a set of edges.
We define $d(v)$ as the degree of node $v$.
Each incident edge of node $v$ is assigned a locally-unique port number in $\{1,\dots ,d(v)\}$.
That is, on node $v$, the port number of edge $(v,u)$ is different from that of edge $(v,w)$ for node $w\neq u$.
Nodes do not have IDs or memories.

\paragraph{Agent}
We denote by $\textit{MA}=\{a_1,a_2,\dots,a_k\}$ the set of $k$ agents.
Each agent $a_i\in \textit{MA}$ has an unique ID denoted by $a_i.\mathit{id}\in \mathbb{N}$ and is equipped with an infinite amount of memory.
Also, agents know the upper bound $N$ of the number of nodes, but they know neither $n$, $k$, nor the IDs of other agents.
Agents cannot mark visited nodes or traversed edges in any way.
An agent is modeled as a state machine $(S,\delta)$, where $S$ is a set of agent states and $\delta$ is a state transition function.
A state is represented by a tuple of the values of all the variables that an agent has.
Each agent has a special state that indicates the termination of an algorithm, called a terminal state.
If an agent transitions into a terminal state, it never moves or updates its state after that.

All agents start an algorithm at the same time, and the initial nodes of the agents are chosen by an adversary.
All agents repeatedly and synchronously execute a round.
In each round, every agent $a_i$ executes the following three operations:
\begin{description}
  \item[\emph{Look}]Agent $a_i$ learns the state of $a_i$, the degree $d(u)$ of the current node $u$, and the port number $i$ of the edge through which the agent arrived at node $u$ (or $a_i$ notices that $u$ is an initial node). Also, if multiple agents exist at node $u$, $a_i$ learns states of all agents at node $u$, including agents in a terminal state. We define $A_i\subseteq \textit{MA}$ as the set of agents existing at node $u$ and $a_i$.
  \item[\emph{Compute}]Agent $a_i$ computes function $\delta$ using the information obtained in the previous Look operation as input. The output is the next agent state, whether it stays or leaves, and the outgoing port number if it leaves.
  \item[\emph{Move}]If $a_i$ decides to stay, it stays at the current node until the beginning of the next round. If $a_i$ decides to leave, it leaves through the decided outgoing port number and arrives at the destination node before the beginning of the next round. 
\end{description}
Note that, if two agents traverse the same edge in different directions at the same time, the agents cannot notice this fact.

\paragraph{Byzantine Agent}
There are $f$ weakly Byzantine agents in the agent system.
Weakly Byzantine agents act arbitrarily apart from an algorithm, except for changing their IDs.
If multiple agents meet Byzantine agents at the same node, all of them learn the same statuses of the Byzantine agents in the Look operation.
We call all agents except weakly Byzantine agents \emph{good} agents and denote by $g=k-f$ the number of good agents.
Good agents know neither the actual number $f$ of Byzantine agents nor the upper bound of $f$.

\subsection{Gathering Problem}
\label{sec:GatheringProblem}
The gathering problem requires all good agents to transition into the terminal state at the same node.
This problem allows agents to enter a terminal state at different times.
We measure the time complexity of a gathering algorithm by the number of rounds required for the last good agent to transition into the terminal state.

\section{Building Blocks}
In this section, we describe two existing algorithms that are used as building blocks to design our proposed algorithm in Section \ref{sec:ByzantineGatheringAlgorithm}.

\subsection{Rendezvous Procedure}
The proposed algorithm uses a rendezvous procedure, which allows two different agents to meet at the same node in any connected graph composed of at most $N$ nodes if each of the agents gives a different ID and $N$ as inputs to the procedure.
The procedure is a well-known combination of an ID transformation procedure and an exploration procedure.
The ID transformation procedure is proposed by Dessmark et al.~\cite{Dessmark2006}.
The exploration procedure is based on universal exploration sequences (UXS) and is a corollary of the result of Ta{-}Shma et al.~\cite{Ta-Shma2014}.
We call this rendezvous procedure $\texttt{REL}(id)$, where $id$ is an ID given as input.
If the execution time of the exploration procedure is $\TIME{EX}$, the procedure from \cite{Dessmark2006} allows two different agents to meet at the same node in at most $(2\lfloor\log (id)\rfloor +6)\TIME{EX}$ rounds.
Also, the exploration procedure used in $\texttt{REL}(id)$ allows an agent to visit all nodes of the same type of graph in at most $\TIME{EX}=O(N^5\log(N))$ if the agent knows that the number of nodes is at most $N$.
Thus, the execution time $\TIME{REL}(id)$ of $\texttt{REL}(id)$ is $O(N^5\log(N)\log(id))$ rounds.
We have the following lemma about $\texttt{REL}(id)$.

\begin{lemma}[\cite{Dessmark2006}]
\label{lem:RendezvousAlgorithm}
Let $a_i$ and $a_j$ be two different agents, and $l_1$ (resp. $l_2$) be ID that $a_i$ (resp. $a_j$) has.
Assume that $a_i$ starts $\texttt{REL}(l_1)$ in round $r_i$, $a_j$ starts $\texttt{REL}(l_2)$ in round $r_j$, and $l_1\neq l_2$ holds.
Then, agents $a_i$ and $a_j$ meet at the same node before round $\max(r_i,r_j)+\TIME{REL}(l_{min})$, where $l_{min}=\min(l_1,l_2)$.
Furthermore, $a_i$ (resp. $a_j$) visits all nodes by round $r_i+\TIME{REL}(a_i.id)$ (resp. round $r_j+\TIME{REL}(a_j.id)$).
\end{lemma}

\noindent For integer $t\geq 0$, we write the procedure of the $t$-th round of $\texttt{REL}(id)$ by $\texttt{REL}(id)(t)$.

\subsection{A Parallel Consensus Algorithm in Byzantine Synchronous Message-Passing Systems}
\label{subsec:PBC}
The proposed algorithm uses a parallel consensus algorithm in \cite{Khanchandani2021} working in Byzantine Synchronous Message-Passing Systems by simulating the algorithm on agents systems, as we describe Section \ref{sec:ByzantineGatheringAlgorithm}.
In this section, we summarize the model and the property of the algorithm.

\paragraph{Model}
A message-passing distributed system, in which processors communicate by sending messages, is modeled by an undirected complete graph with $m$ nodes.
The nodes have unique IDs, and these IDs do not necessarily have to be consecutive.
The system includes at most $b$ Byzantine nodes, which can act arbitrarily apart from an algorithm.
We call all nodes except Byzantine nodes \emph{good} nodes.
At the beginning of an execution, each node knows its ID only and knows neither the number $m$ of nodes, the number $b$ of Byzantine nodes, nor the other nodes' IDs.
The system is synchronous, that is, nodes repeat synchronous phases.
In a phase $p$, every good node executes local computation, sends messages to some nodes, and then receives the messages that were sent to it in phase $p$.
Node $v$ can send a message $\textit{msg}$ in two ways: (1) $v$ broadcasts $\textit{msg}$ to all nodes, or (2) $v$ sends $\textit{msg}$ to a specific node that $v$ knows its ID.
Every message has the ID of its sender; thus, when a node receives a message, it can obtain the sender's ID.
There is no restriction on the actions of Byzantine nodes except for falsifying their IDs to a directly communicating node.

\paragraph{Parallel Byzantine Consensus Problem}
\label{subsec:PBCProblem}
Each good node $v$ has a set $S_v$ composed of $k_v$ input pairs $(id_v^i,x_v^i)$ $(1\leq i\leq k_v)$, where $id_v^i$ is an ID of the input pair and $x_v^i$ is an input number.
We say an algorithm solves the parallel Byzantine consensus problem if, when each good node $v$ starts with set $S_v$ as an input, each node outputs a set of pairs subject to the following conditions:

\begin{description}
  \item[Validity 1] If $(id,x)$ is an input pair of every good node and $x\neq \bot$, then the output set of a good node must include $(id,x)$.
  \item[Validity 2] If $(id,x)$ is not an input pair of any good node, then the output set of any good node does not include $(id,x)$.
  \item[Agreement] If the output set of a good node includes $(id,x)$, then the output sets of all other good nodes must include $(id,x)$ as well.
  \item[Termination] Every good node outputs a set of pairs exactly once in a finite number of phases.
\end{description}

\noindent If an algorithm satisfies the above four conditions, we say it satisfies the parallel Byzantine consensus property (in short, PBC property).
The PBC property allows that, if $(id,x)$ is included in an input set of a part of good nodes, but not all good nodes, $(id,x)$ may not included in the output set of any good node.

\paragraph{Parallel Byzantine Consensus Algorithm}
The proposed algorithm uses the parallel Byzantine consensus algorithm in \cite{Khanchandani2021}.
We call this algorithm $\texttt{PCONS}(S)$, where $S$ is a set given as input.
We have the following lemma about $\texttt{PCONS}(S)$.

\begin{lemma}[\cite{Khanchandani2021}]
\label{lem:PBCAlgorithm}
Assume that more than $3b$ nodes exist in a system.
If every good node $v$ simultaneously starts $\texttt{PCONS}(S_v)$ with a set $S_v$ as input, its execution satisfies the PBC property.
Every good node outputs a set in $O(b)$ phases, and its output time differs by at most one phase among good nodes.
\end{lemma}

\section{Byzantine Gathering Algorithm}
\label{sec:ByzantineGatheringAlgorithm}
To achieve the gathering, the proposed algorithm uses a subroutine that is used as a building block.
In this section, we first explain the overview of the proposed algorithm.
After that, we give an idea and a detailed description of the subroutine, and then we show an algorithm to achieve the gathering.
Throughout the paper, we assume $k=g+f\geq 9f+8$, which implies that there are at least $8f+8$ good agents in the network.
Recall that agents know $N$, but do not know $n$, $k$, or $f$.

\subsection{Overview}
\label{sec:Overview}

Here, we give the overview of the proposed Byzantine gathering algorithm, which aims to gather all good agents at a single node.
For simplicity, we assume that agents know $f$ here, and will remove this assumption in Sections \ref{sec:MakeReliableGroupConstruction} and \ref{sec:GatheringAlgorithm}.
The underlying idea of the algorithm is made of the following three steps:
\begin{itemize}
  \item[(1)] All agents collect all agent IDs by using the rendezvous algorithm $\texttt{REL}$.
  \item[(2)] After collecting all good agents' IDs, every agent decides on an ID that is common to other good agents as a target ID from the collected IDs.
  \item[(3)] An agent $a_{target}$ with the target ID stays at the current node and the other agents search for $a_{target}$ using $\texttt{REL}$.
\end{itemize}
If there are no Byzantine agents, all agents can decide on a common target ID by choosing the smallest ID of the collected IDs.
Therefore, in Step (3), all agents gather at the node where the agent with the smallest ID exists.
However, if there is a Byzantine agent, that idea fails.
Let us consider the case where a Byzantine agent $\textit{Byz}\in \textit{MA}$ has the smallest ID.
If $\textit{Byz}$ meets only a part of good agents in Step (1), the other good agents do not choose $\textit{Byz}.id$ as a target ID.
Therefore, good agents are divided into two or more groups.
Also, even if all agents know $\textit{Byz}.id$, $\textit{Byz}$ can avoid meeting the other agents in Step (3), and all good agents keep searching endlessly for $a_{target}$.

To solve these problems, the proposed algorithm suppresses the influence of Byzantine agents by letting several agents create a reliable group such that good agents can trust the behavior of the group.
After collecting all good agents' IDs, agents execute the following three steps:
\begin{itemize}
  \item[(a)] Agents create a group candidate of at least $3f+1$ agents.
  \item[(b)] Agents in the group candidate make a common ID set by using the parallel Byzantine consensus algorithm in Section \ref{subsec:PBC}.
  \item[(c)] By using the common ID set, agents in the group candidate gather to create a reliable group composed of at least $f+1$ good agents.
\end{itemize}
The goal of Steps (a) and (b) is to make at least $3f+1$ agents make a common ID set.
To do this, we use the parallel Byzantine consensus algorithm in Section \ref{subsec:PBC}.
Since the consensus algorithm assumes message-passing systems, agents simulate the system by using the rendezvous algorithm $\texttt{REL}$.
Simply put, agents exchange messages when they meet other agents by $\texttt{REL}$.
In Step (a), agents create a group candidate of at least $3f+1$ agents such that they can send messages to each other among the group candidate.
In Step (b), as an input of the consensus algorithm, each agent uses the set of agent IDs (known to the agent) in the same group candidate.
If the group candidate is composed of at least $3f+1$ agents, the output is common and includes IDs of all good agents in the group candidate.
In Step (c), agents in a group candidate decide on a target ID based on the common ID set and gather at the node where an agent with the target ID exists.
If at least $2f+1$ agents gather, they create a reliable group composed of at least $f+1$ good agents.
If agents do not gather sufficiently, the agents determine the next target ID and find the agent with the new target ID.
The algorithm ensures that all good agents in the group candidate eventually gather and create a reliable group.

Once at least one reliable group is created, the proposed algorithm can achieve the gathering as follows.
Good agents in the reliable group decide on the smallest agent ID in the group as a group ID and execute $\texttt{REL}$ using the group ID.
On the other hand, good agents not in the reliable group execute $\texttt{REL}$ using their own IDs.
Furthermore, when each good agent meets the reliable group with a smaller group ID, it accompanies the group.
All the group agents in the reliable group are at the same node, act identically, and have the same group ID.
If a good agent meets the reliable group, the agent trusts the group since the group consists of at least $f+1$ agents with the same group ID, which implies that the group contains at least one good agent.
As a result, all good agents accompany the reliable group with the smallest ID and achieve the gathering.

The gathering algorithm by Hirose et al.~\cite{Hirose2021} employs another strategy to create a reliable group from collecting IDs instead of using a consensus algorithm.
In the algorithm, each good agent searches for one of the agents with the smallest $f+1$ IDs of the collected IDs to gather at the node where the agent is.
Because good agents may be divided into $\Omega(f)$ nodes in this strategy, this strategy requires at least $\Omega(f^2)$ good agents to guarantee that a reliable group is created at any of those nodes.
On the other hand, the proposed algorithm uses the strategy such that $\Omega(f)$ good agents make a common ID set, search for a target agent one by one synchronously, and try to gather at the node with any of $\Omega(f)$ agents.
Therefore, the algorithm requires $\Omega(f)$ good agents, i.e., the key to the algorithm is the reliable group creation procedure using the consensus algorithm.

\subsection{Algorithm to Create a Reliable Group}
\label{sec:MakeReliableGroupConstruction}
In this section, we explain an algorithm to create a reliable group, called $\texttt{MakeReliableGroup}$, by assuming that $k=g+f\geq 9f+8$.
Recall that agents know $N$, but do not know $n$, $k$, or $f$.

\subsubsection{Idea of the Algorithm}
\label{subsec:MRGC_Idea}
As mentioned in Section \ref{sec:Overview}, in $\texttt{MakeReliableGroup}$, agents in the group candidate make a common ID set and search for agents with target IDs.
Algorithm $\texttt{MakeReliableGroup}$ uses the parallel Byzantine consensus algorithm $\texttt{PCONS}$ for at least $2f+1$ good agents to have a common ID set.
However, since $\texttt{PCONS}$ assumes a Byzantine synchronous message-passing model, we cannot use $\texttt{PCONS}$ directly.
Therefore, $\texttt{MakeReliableGroup}$ simulates the model on an agent system as follows.

First, $\texttt{MakeReliableGroup}$ selects a group candidate composed of at least $3f+1$ agents and finds time $T$ that is sufficiently long to meet all the agents in the group candidate by $\texttt{REL}$.
Then, agents regard the time interval of $T$ rounds as a phase in the message-passing model, and simulate the behavior of the phase during the $T$ rounds.
More concretely, each agent $a_i$ in the group candidate executes $\texttt{REL}(a_i.id)$ for $T$ rounds and, if $a_i$ meets another agent $a_j$ in the group candidate, $a_i$ shares a message that $a_i$ sent in the previous phase with $a_j$.
Then, in the last round of the interval, $a_i$ executes the computation of the phase using the messages collected by the current phase.
Since $a_i$ can meet all good agents in the group candidate during the $T$ rounds, this behavior can simulate the broadcast of messages in the Byzantine synchronous message-passing model.

This simulation requires agents to (1) construct a group candidate composed of at least $3f+1$ agents, (2) start $\texttt{PCONS}$ simultaneously with the agents in the same group candidate, and (3) know time $T$ that is long enough to meet the agents in the same group candidate.
To meet the requirements, agents synchronously repeat a \emph{cycle}.
Each agent sets the length of the first cycle as $T_{ini}$ rounds, where $T_{ini}$ is a given positive integer, and doubles the length every cycle, i.e., that of the second cycle is $2\cdot T_{ini}$ rounds, that of the third cycle is $4\cdot T_{ini}$ rounds, and so on.
If the length of a cycle is long enough for agent $a_i$ to meet all other good agents by $\texttt{REL}(a_i.id)$, $a_i$ starts $\texttt{REL}(a_i.id)$ for the cycle.
If the length of a cycle is long enough for at least $3f+1$ agents to meet each other, $\texttt{MakeReliableGroup}$ regards them as a group candidate and makes them start $\texttt{PCONS}$ simultaneously.
By this behavior, $\texttt{MakeReliableGroup}$ can achieve both (1) and (2).
Furthermore, since the length of the cycle is long enough for at least $3f+1$ agents to meet each other, $\texttt{MakeReliableGroup}$ can achieve (3) by defining $T$ as the length of the cycle when the agents start $\texttt{PCONS}$.

Algorithm $\texttt{MakeReliableGroup}$ consists of four stages: $\textit{CollectID}$, $\textit{MakeCandidate}$, $\textit{AgreeID}$, and $\textit{MakeGroup}$ stages.
In the $\textit{CollectID}$ stage, agents collect IDs of all good agents.
In the $\textit{MakeCandidate}$ stage, agents select the group candidate.
In the $\textit{AgreeID}$ stage, agents in the group candidate obtain a common ID set by using consensus algorithm $\texttt{PCONS}$.
In the $\textit{MakeGroup}$ stage, agents create a reliable group.

\subsubsection{Details of the Algorithm}
\begin{algorithm}[t]
  \caption{$\texttt{MakeReliableGroup}$}
  \label{alg:MakeReliableGroupConstruction}
  \begin{algorithmic}[1]
    \State $a_i.\textit{elapsed}\leftarrow a_i.\textit{elapsed}+1$
    \If{$a_i.\textit{stage}=\textit{CollectID}$}
      \State // While executing $\texttt{CollectIDStage}$, $a_i$ executes $a_i.\textit{length}\leftarrow 2\cdot a_i.\textit{length}$.
      \State Execute $\texttt{CollectIDStage}$
    \ElsIf{$a_i.\textit{stage}=\textit{MakeCandidate}$}
      \State // While executing $\texttt{MakeCandidateStage}$, $a_i$ executes $a_i.\textit{length}\leftarrow 2\cdot a_i.\textit{length}$.
      \State Execute $\texttt{MakeCandidateStage}$
    \ElsIf{$a_i.\textit{stage}=\textit{AgreeID}$}
      \State // While executing $\texttt{AgreeIDStage}$, $a_i$ executes $a_i.\textit{count}\leftarrow a_i.\textit{count}+1$
      \State Execute $\texttt{AgreeIDStage}$
    \ElsIf{$a_i.\textit{stage}=\textit{MakeGroup}$}
      \State // While executing $\texttt{MakeGroupStage}$, $a_i$ executes $a_i.\textit{count}\leftarrow a_i.\textit{count}+1$
      \State Execute $\texttt{MakeGroupStage}$
    \EndIf
  \end{algorithmic} 
\end{algorithm}

\begin{table}[t]
  \centering
  \caption{Variables of agent $a_i$.}
  \label{tab:variable}
  \begin{tabular}{c|c|p{0.6\hsize}}\hline
    Variable & Initial value & \multicolumn{1}{c}{Explanation}\\ \hline
    $\textit{stage}$ & $\textit{CollectID}$ & The current stage of $a_i$. This variable takes one of the following values: $\textit{CollectID}$, $\textit{MakeCandidate}$, $\textit{AgreeID}$, $\textit{MakeGroup}$. \\ \hline
    $\textit{length}$ & $T_{ini}$ & The length of the current cycle. \\ \hline
    $\textit{elapsed}$ & $0$ & The number of rounds from the beginning of the current cycle. \\ \hline
    $\textit{count}$ & $0$ & The number of cycles from the beginning of the $\textit{AgreeID}$ stage. \\ \hline
    $\textit{ready}$ & $\textit{False}$ & $\textit{True}$ if and only if $a_i$ has met a certain condition, which represents the decision that $a_i$ is ready to transition into the $\textit{AgreeID}$ stage in the $\textit{MakeCandidate}$ stage. \\ \hline
    $\textit{endMakeCandidate}$ & $\textit{False}$ & $\textit{True}$ if and only if $a_i$ has met the condition to transition into the $\textit{AgreeID}$ stage. \\ \hline
    $\textit{gid}$ & $\infty$ & The group ID of the reliable group to which $a_i$ belongs. \\ \hline
    $R$ & $\emptyset$ & A set of IDs of agents such that $a_i$ knows they satisfy $\textit{ready}=\textit{True}$. \\ \hline
    $S_p$ & $\{a_i.id\}$ & A set of agent IDs that $a_i$ has collected in the $\textit{CollectID}$ stage. \\ \hline
    $S_c$ & $\emptyset$ & An output of $\texttt{PCONS}(S_p)$. \\ \hline
    $P_p$ & $\emptyset$ & A set of IDs of agents such that $a_i$ knows they belong to the same group candidate as $a_i$. \\ \hline
    $P_c$ & $\emptyset$ & An output of $\texttt{PCONS}(P_p)$, which is used as a common ID set. Its elements are ordered in increasing order. \\ \hline
  \end{tabular}
\end{table}

\begin{figure}[t]
  \centering
  \begin{tabular}{c}
    \begin{minipage}[t]{0.9\hsize}
      \centering
      \includegraphics[width=\textwidth]{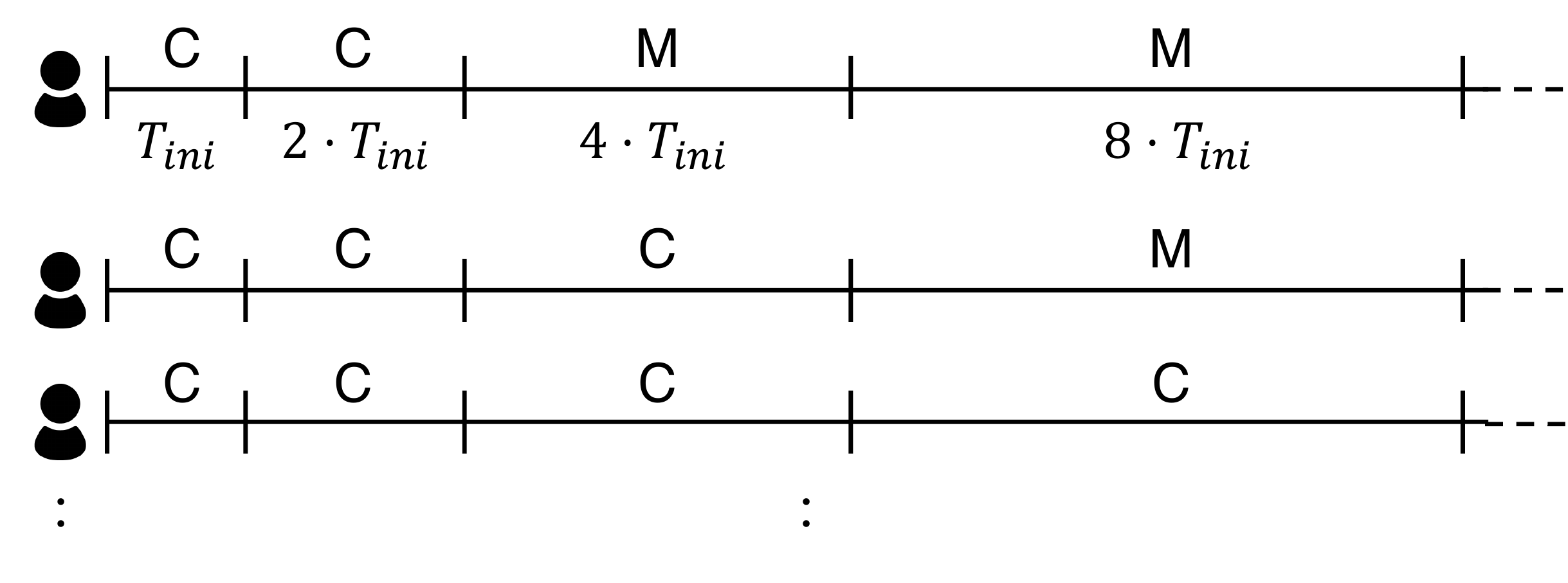}
      \subcaption{At starting \texttt{MakeReliableGroup}}
      \label{fig:StageFlowMulti_a}
    \end{minipage} \\\\
    \begin{minipage}[t]{0.9\hsize}
      \centering
      \includegraphics[width=\textwidth]{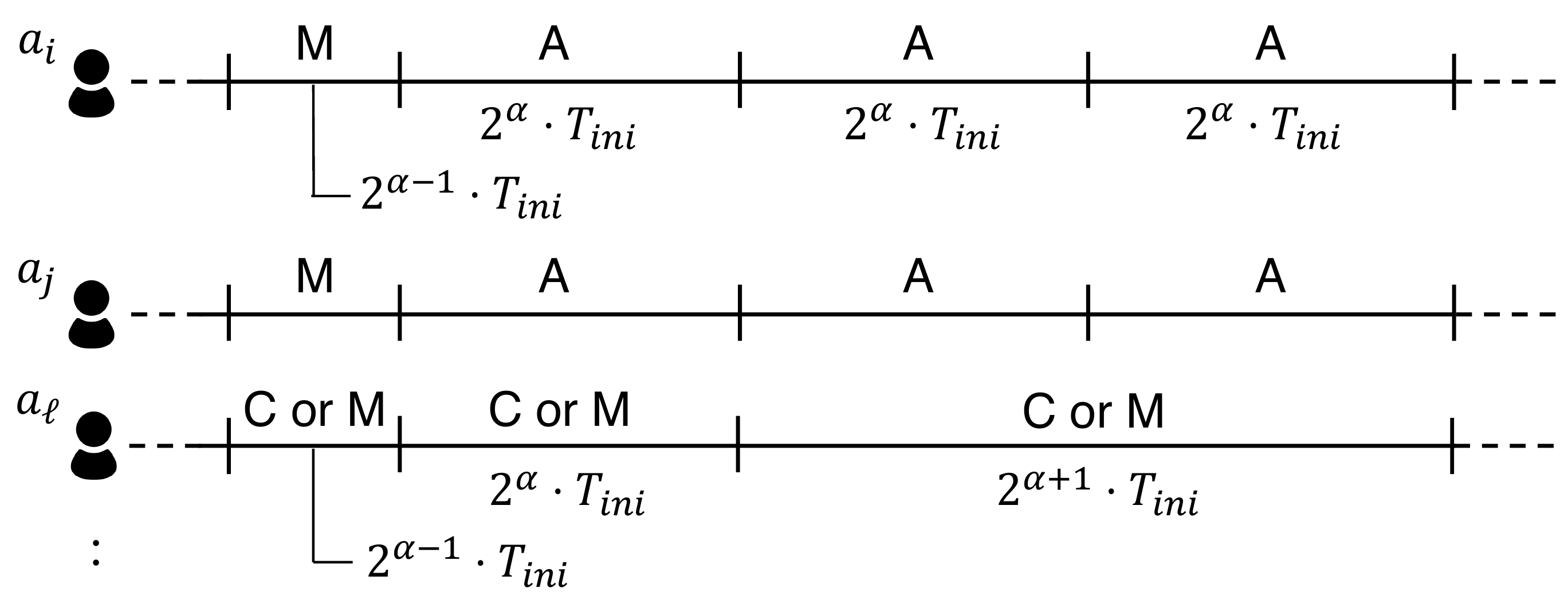}
      \subcaption{At starting the $\textit{AgreeID}$ stage}
      \label{fig:StageFlowMulti_b}
    \end{minipage}
  \end{tabular}
  \caption{The stage flow of Algorithm $\texttt{MakeReliableGroup}$. Notions C, M, and A represent cycles of the $\textit{CollectID}$ stage, $\textit{MakeCandidate}$ stage, and the $\textit{AgreeID}$ stage, respectively.}
  \label{fig:StageFlowMulti}
\end{figure}

Algorithm \ref{alg:MakeReliableGroupConstruction} shows the behavior of each round of Algorithm $\texttt{MakeReliableGroup}$ and executes one of Algorithms \ref{alg:CollectIDStage}--\ref{alg:MakeGroupStage} depending on the current stage.
In $\texttt{MakeReliableGroup}$, the procedure $\texttt{WAIT}()$ means that an agent stays at the current node for one round.
Table \ref{tab:variable} summarizes variables used in $\texttt{MakeReliableGroup}$.
Variable $\textit{stage}$ keeps the current stage of $a_i$ and its initial value is $\textit{CollectID}$.
Agent $a_i$ has variable $\textit{length}$ to keep the length of the current cycle
Agent $a_i$ doubles $\textit{length}$ at the last of each cycle if $a_i.\textit{stage}\in \{\textit{CollectID},\textit{MakeCandidate}\}$.
The initial value of $\textit{length}$ is $T_{ini}$, where $T_{ini}$ is a given positive integer.
Agent $a_i$ has variable $\textit{elapsed}$ to keep the number of rounds from the beginning of the current cycle.
Variable $\textit{count}$ maintains the number of cycles that elapsed after the beginning of the $\textit{AgreeID}$ stage.

The overall flow of $\texttt{MakeReliableGroup}$ is shown in Fig.~\ref{fig:StageFlowMulti}.
In $\texttt{MakeReliableGroup}$, $a_i$ executes the $\textit{CollectID}$ stage, the $\textit{MakeCandidate}$ stage, the $\textit{AgreeID}$ stage, and the $\textit{MakeGroup}$ stage in this order (Algorithm \ref{alg:MakeReliableGroupConstruction}). 
All good agents have the same initial length of a cycle and double their length every cycle until they start the $\textit{AgreeID}$ stage.
However, as we explain later, good agents may transition into the next stage at different cycles.
Therefore, good agents have the same length of a cycle until they start the $\textit{AgreeID}$ stage.
The following observation shows this fact formally.
For simplicity, we denote $\gamma$-th cycle of an agent $a_i$ and its length by $c_i^\gamma$ and $|c_i^\gamma|$, respectively.

\begin{observation}
\label{obs:XsOfCollectIDandPreparationAreSame}
Let $a_i$ and $a_j$ be good agents.
If $a_i$ and $a_j$ are in the $\textit{CollectID}$ stage or the $\textit{MakeCandidate}$ stage, $|c_i^\gamma|=|c_j^\gamma|$ holds for any $\gamma$ and they started these cycles at the same time.
\end{observation}

In $\texttt{MakeReliableGroup}$, agents that start the $\textit{AgreeID}$ stage simultaneously compose a group candidate.
An agent $a_i$ can check whether another agent belongs to the same group candidate as follows.
Assume that $a_i$ stores $\textit{AgreeID}$ in $a_i.\textit{stage}$ in round $r$.

Let $a_j$ be an agent that also stores $\textit{AgreeID}$ in $a_j.\textit{stage}$ in round $r$.
By Observation \ref{obs:XsOfCollectIDandPreparationAreSame}, $|c_i^\gamma|=|c_j^\gamma|$ holds for $c_i^\gamma$ and $c_j^\gamma$ that include round $r$.
Furthermore, since an agent does not double the length of the cycle in the $\textit{AgreeID}$ and the $\textit{MakeGroup}$ stages, $|c_i^\varepsilon|=|c_j^\varepsilon|$ holds for any $\varepsilon>\gamma$.

Let us consider another agent $a_\ell$ that stores $\textit{AgreeID}$ in $a_\ell.\textit{stage}$ in round $r'$ ($r'\neq r$).
Since the total number of cycles that $a_\ell$ has executed in the $\textit{CollectID}$ and the $\textit{MakeCandidate}$ stages is greater or less than that of $a_i$, the number of updates of $a_\ell.\textit{length}$ is different from that of $a_i.\textit{length}$.
Thus, $|c_i^\varepsilon|=|c_\ell^\zeta|$ does not hold, where $c_\ell^\zeta$ is the cycle when $a_\ell$ starts the $\textit{AgreeID}$ stage.
Hence, when $a_i$ witnesses $a_\ell$ at the current node after it starts the $\textit{AgreeID}$ stage, $a_i$ can understand that $a_\ell$ started the $\textit{AgreeID}$ stage at a different round by observing $a_\ell.\textit{length}$.
In other words, $a_i$ can understand that $a_\ell$ is a member of the group candidate different from $a_i$.
The following observation summarizes this discussion.

\begin{observation}
\label{obs:CountsOfConsensusAndGathringAreSame}
Let $a_i$ and $a_j$ be good agents that have entered the $\textit{AgreeID}$ or the $\textit{MakeGroup}$ stage.
If $a_i$ and $a_j$ started the $\textit{AgreeID}$ stage at the same time, $|c_i^\gamma|=|c_j^\gamma|$ holds for any $\gamma$, otherwise $|c_i^\gamma|=|c_j^\gamma|$ does not hold.
\end{observation}

Consider the case where $a_i$ starts the $\textit{AgreeID}$ stage faster than $a_\ell$ like Fig.~\ref{fig:StageFlowMulti_b}.
Since $a_i$ does not update $a_i.\textit{length}$ anymore, the length of each cycle of the $\textit{AgreeID}$ and the $\textit{MakeGroup}$ stages is identical.
On the other hand, $a_\ell$ doubles $a_\ell.\textit{length}$ every cycle until it starts the $\textit{AgreeID}$ stage.
This implies that $a_\ell.\textit{length}\equiv 0\ (\bmod\ a_i.\textit{length})$ holds.
That is, when $a_\ell$ starts a cycle, $a_i$ also starts a cycle.
From this discussion and Observations \ref{obs:XsOfCollectIDandPreparationAreSame} and \ref{obs:CountsOfConsensusAndGathringAreSame}, we have the following observation.

\begin{observation}
\label{obs:LargerXGoodStartTimeAndSmallerXGoodStartTimeSame}
Let $a_i$ and $a_j$ be good agents such that $a_j$ starts the $\textit{AgreeID}$ stage no earlier than $a_i$.
When $a_j$ starts a cycle, $a_i$ also starts a cycle at the same time.
\end{observation}

\paragraph{$\textit{CollectID}$ Stage}
\begin{algorithm}[t]
  \caption{$\texttt{CollectIDStage}$}
  \label{alg:CollectIDStage}
  \begin{algorithmic}[1]
    \State $a_i.R\leftarrow a_i.R\cup \{a_j.id\mid a_j\in A_i \wedge a_j.\textit{ready}=\textit{True}\}$
    \If{$2\cdot (\TIME{REL}(a_i.id)+1)>a_i.\textit{length}$}
      \If{$a_i.\textit{length}=a_i.\textit{elapsed}$}
        \State $a_i.\textit{elapsed}\leftarrow 0$
        \State $a_i.\textit{length}\leftarrow 2\cdot a_i.\textit{length}$
      \EndIf
      \State Execute $\texttt{WAIT}()$
    \Else \Comment{$2\cdot (\TIME{REL}(a_i.id)+1)\leq a_i.\textit{length}$}
      \State $a_i.S_p\leftarrow a_i.S_p\cup \{a_j.id \mid a_j \in A_i\}$
      \If{$a_i.\textit{length}>a_i.\textit{elapsed}$}
        \State Execute $\texttt{REL}(a_i.id)(a_i.\textit{elapsed})$
      \Else
        \State $a_i.\textit{elapsed}\leftarrow 0$
        \State $a_i.\textit{length}\leftarrow 2\cdot a_i.\textit{length}$
        \State $a_i.\textit{stage}\leftarrow \textit{MakeCandidate}$
        \State Execute $\texttt{WAIT}()$
      \EndIf
    \EndIf
  \end{algorithmic} 
\end{algorithm}

Algorithm \ref{alg:CollectIDStage} is the pseudo-code of the $\textit{CollectID}$ stage.
This stage aims to collect IDs of all good agents.
An agent $a_i$ uses variable $S_p$ to record the IDs collected in the $\textit{CollectID}$ stage.

If $a_i$ witnesses an agent $a_j$ with $a_j.\textit{ready}=\textit{True}$ at the current node at the beginning of a round, $a_i$ stores $a_j.id$ in variable $a_i.R$ to record IDs of such agents (Line 1 of Algorithm \ref{alg:CollectIDStage}).
Recall that $A_i$ is a set of agents (including $a_i$) that stay at the current node of $a_i$ at the beginning of a round.
We will explain the details of variable $a_i.\textit{ready}$ and $a_i.R$ in Section \ref{subsubsec:MakeCandidatestage}.

If $a_i.\textit{length}<2\cdot (\TIME{REL}(a_i.id)+1)$ holds in a cycle, $a_i$ stays at the current node for the cycle.
Additionally, $a_i$ updates variables at the last round of the current cycle for the next cycle.
To determine whether the current round is the last one of the current cycle, $a_i$ uses variable $\textit{elapsed}$, which keeps the number of rounds elapsed during the current cycle.
If $a_i.\textit{length}=a_i.\textit{elapsed}$ holds, $a_i$ understands that the current round is the last one of the current cycle and updates variables $a_i.\textit{elapsed}$ and $a_i.\textit{length}$ for the next cycle (Lines 3--6).

If $a_i.\textit{length}\geq 2\cdot (\TIME{REL}(a_i.id)+1)$ holds in a cycle, $a_i$ collects IDs of all good agents using $\texttt{REL}(a_i.id)$ (Lines 8--18).
Agent $a_i$ uses variable $S_p$ to record collected agent IDs.
If $a_i$ witnesses $a_j$ at the current node at the beginning of a round, $a_i$ stores $a_j.id$ in $a_i.S_p$ (Line 9).
After that, if the current round is not the last round, that is, $a_i.\textit{length}>a_i.\textit{elapsed}$ holds, $a_i$ executes $\texttt{REL}(a_i.id)(a_i.\textit{elapsed})$ (Lines 10--11).
Otherwise, that is, if $a_i.\textit{length}=a_i.\textit{elapsed}$ holds, $a_i$ stores $\textit{MakeCandidate}$ in $a_i.\textit{stage}$ (to move the $\textit{MakeCandidate}$ stage in the next cycle) and stays at the current node for one round (Lines 12--17).
As we prove later, if $a_i.\textit{length}\geq 2\cdot (\TIME{REL}(a_i.id)+1)$ holds in the current cycle, $a_i$ meets all other good agents by the end of the last round of the current cycle.
Therefore, when $a_i$ finishes the last round of that cycle, $a_i.S_p$ includes IDs of all good agents.

\paragraph{$\textit{MakeCandidate}$ Stage}
\label{subsubsec:MakeCandidatestage}
\begin{algorithm}[t]
  \caption{$\texttt{MakeCandidateStage}$}
  \label{alg:MakeCandidateStage}
  \begin{algorithmic}[1]
    \State $a_i.R\leftarrow a_i.R\cup \{a_j.id\mid a_j\in A_i \wedge a_j.\textit{ready}=\textit{True}\}$
    \If{$a_i.\textit{elapsed}=1\wedge (\textrm{either (1) or (2) holds}) \wedge a_i.\textit{ready}=\textit{False}$}
      \State // (1) $|\{id\in a_i.S_p\mid a_i.\textit{length}\geq 4\cdot (\TIME{REL}(id)+1)\}|\geq(8/9)|a_i.S_p|$
      \State // (2) $|a_i.R|\geq (4/9)|a_i.S_p|$
      \State $a_i.\textit{ready}\leftarrow\textit{True}$
      \State $a_i.R\leftarrow a_i.R\cup \{a_i.id\}$
    \EndIf
    \If{$a_i.\textit{elapsed}=1\wedge |a_i.R|\geq (6/9)|a_i.S_p|$}
      \State $a_i.\textit{endMakeCandidate}\leftarrow \textit{True}$
    \EndIf
    \If{$a_i.\textit{length}>a_i.\textit{elapsed}$}
      \State Execute $\texttt{REL}(a_i.id)(a_i.\textit{elapsed})$
    \Else
      \State $a_i.\textit{elapsed}\leftarrow 0$
      \State $a_i.\textit{length}\leftarrow 2\cdot a_i.\textit{length}$
      \If{$a_i.\textit{endMakeCandidate}=\textit{True}$}
        \State $a_i.\textit{stage}\leftarrow \textit{AgreeID}$
      \EndIf
      \State Execute $\texttt{WAIT}()$
    \EndIf
  \end{algorithmic} 
\end{algorithm}

Algorithm \ref{alg:MakeCandidateStage} is the pseudo-code of the $\textit{MakeCandidate}$ stage.
This stage aims to create a group candidate consisting of at least $3f+1$ good agents.
As we explain in Section \ref{subsubsec:MakeGroupStage}, good agents estimate $f$ to create a reliable group consists of at least $2f+1$ agents.
However, their estimated values differ by at most $f$.
Thus, $\texttt{MakeReliableGroup}$ requires at least $3f+1$ good agents to allow for that error.
In order to describe the $\textit{MakeCandidate}$ stage clearly, we define the group candidate as follows.

\begin{definition}[Group candidate]
\label{def:GroupCandidate}
A set $\textit{GC}$ of agents is a group candidate if and only if all good agents in $\textit{GC}$ satisfy the following:
(1) All the good agents in $\textit{GC}$ start the $\textit{AgreeID}$ stage at the same time.
(2) For any good agent $a_i\in \textit{GC}$, $a_i.\textit{length}\geq 4\cdot (\TIME{REL}(a_i.id) +1)$ holds after the beginning of the $\textit{MakeCandidate}$ stage.
\end{definition}

By the behavior of the $\textit{CollectID}$ stage, when an agent $a_i$ starts the $\textit{MakeCandidate}$ stage, the length of $a_i$'s cycle is at least $4\cdot (\TIME{REL}(a_i.id) +1)$.
Furthermore, since the $\textit{MakeCandidate}$ stage does not include the action of reducing the length of a cycle, $a_i$ already satisfies Requirement (2) of Definition \ref{def:GroupCandidate}.
Also, by Observation \ref{obs:XsOfCollectIDandPreparationAreSame}, every good agent in the $\textit{MakeCandidate}$ stage has the same length of a cycle.
Therefore, agents that start the $\textit{AgreeID}$ stage at the same time satisfy Requirement (1).
Consequently, if at least $3f+1$ of the good agents start the $\textit{AgreeID}$ stage simultaneously, they achieve the purpose of the $\textit{MakeCandidate}$ stage.

Agent $a_i$ executes $\texttt{REL}(a_i.id)$ every cycle of the $\textit{MakeCandidate}$ stage.
First, if $a_i$ witnesses an agent $a_j$ with $a_j.\textit{ready}=\textit{True}$ at the current node at the beginning of a round, it stores $a_j.id$ in $a_i.R$ (Line 1 of Algorithm \ref{alg:MakeCandidateStage}).
Then, if $a_i$ satisfies either of the following two conditions, $a_i$ stores $\textit{True}$ in $\textit{ready}$ (Lines 2--7).
\begin{itemize}
  \item[(1)] Variable $a_i.S_p$ contains at least $(8/9)|a_i.S_p|$ IDs of agents that have started the $\textit{MakeCandidate}$ stage.
  \item[(2)] Agent $a_i$ has witnessed at least $(4/9)|a_i.S_p|$ agents with $\textit{ready}=\textit{True}$ from the beginning of $\texttt{MakeReliableGroup}$.
\end{itemize}
By the behavior of the $\textit{CollectID}$ stage, for a good agent $a_i$, when the length of $a_j$'s cycle becomes at least $4\cdot (\TIME{REL}(a_j.id) +1)$, $a_j$ starts the $\textit{MakeCandidate}$ stage.
Therefore, $a_i$ can determine Condition (1) by checking whether $a_i.S_p$ contains at least $(8/9)|a_i.S_p|$ IDs, each $id$ of which satisfies $a_i.\textit{length}\geq 4\cdot (\TIME{REL}(id)+1)$.
Also, $a_i$ can decide Condition (2) by judging whether $a_i.R$ includes at least $(4/9)|a_i.S_p|$ IDs.
Next, if $a_i.R$ contains at least $(6/9)|a_i.S_p|$ IDs, $a_i$ stores $\textit{True}$ in variable $\textit{endMakeCandidate}$ (Lines 8--10).
It means that $a_i$ starts the $\textit{AgreeID}$ stage from next cycle.
Finally, if the current round is not the last one of the current cycle, that is, $a_i.\textit{length}>a_i.\textit{elapsed}$ holds, then $a_i$ executes $\texttt{REL}(a_i.id)(a_i.\textit{elapsed})$ to inform other agents of the current state of $a_i.\textit{ready}$ (Lines 11--12).
Otherwise, that is, if $a_i.\textit{length}=a_i.\textit{elapsed}$ holds, $a_i$ stays at the current node for one round (Lines 13--20).
In this case, if $a_i.\textit{endMakeCandidate}=\textit{True}$ holds, $a_i$ stores $\textit{AgreeID}$ in $a_i.\textit{stage}$ before it stays for one round (Lines 16--18).

By the above behavior, at least one group candidate consisting of at least $3f+1$ good agents is created.
The reason is as follows.
Assume that agent $a_{ini}$ is the first good agent that stores $\textit{True}$ in $\textit{endMakeCandidate}$, and let $c_{ini}^\gamma$ be the cycle in which $a_{ini}$ does so.
In this case, the following two situations occur.
\begin{itemize}
  \item[(a)] Each good agent that started the $\textit{MakeCandidate}$ stage has witnessed at least $(6/9)|a_{ini}.S_p|-f$ good agents with $\textit{ready}=\textit{True}$ by the beginning of cycle $c_{ini}^\gamma$.
  \item[(b)] There exists at least one good agent that satisfies Condition (1) (Line 3 of Algorithm \ref{alg:MakeCandidateStage}) by the beginning of cycle $c_{ini}^\gamma$ (proved in Lemma \ref{lem:AgentsSendMessageBySatisfyingAlgorithmMakeCandidate4}).
\end{itemize}
Let $a_\ell$ be a good agent that satisfies Condition (1) by the beginning of cycle $c_{ini}^\gamma$.
By Condition (1) and Situation (b), the following situation occurs: 
\begin{itemize}
    \item[(b')] At least $(8/9)|a_\ell.S_p|-f$ good agents have started the $\textit{MakeCandidate}$ stage by cycle $c_{ini}^\gamma$.
\end{itemize}
Here, 
$(8/9)|a_\ell.S_p|-f\geq (8/9)g-f = (6/9)g+(2/9)g-f \geq (6/9)g+(2/9)(8f+8)-f = (6/9)g+(7/9)f+16/9 > (6/9)k$ holds
because $k\geq |a_\ell.S_p|\geq g$ holds, which we prove by Corollary \ref{cor:GoodKnowAllGoodIds}.
That is, at least $(6/9)k$ good agents have started the $\textit{MakeCandidate}$ stage by cycle $c_{ini}^\gamma$.
Here, let $a_m$ be such a good agent.
By Situation (a), $a_m$ has witnessed at least 
$(6/9)|a_{ini}.S_p|-f$ $\geq (6/9)g-f=(4/9)g+(2/9)g-f\geq (4/9)g+(2/9)(8f+8)-f=(4/9)g+(7/9)f+16/9>(4/9)k\geq (4/9)|a_m.S_p|$ agents with $\textit{ready}=\textit{True}$ before cycle $c_{ini}^\gamma$.
Therefore, since $a_m$ stores $\textit{True}$ in $a_m.\textit{ready}$ before cycle $c_{ini}^\gamma$, $a_m$ witnesses at least $(6/9)k\geq (6/9)|a_m.S_p|$ agents with $\textit{ready}=\textit{True}$ during cycle $c_{ini}^\gamma$.
Thus, at least $(6/9)k\geq (6/9)(9f+8)>6f+1$ good agents store $\textit{True}$ in $\textit{endMakeCandidate}$ in the first round of either $c_{ini}^\gamma$ or $c_{ini}^{\gamma+1}$.
Hence, at least $3f+1$ good agents start the $\textit{AgreeID}$ stage at the same time.

\paragraph{$\textit{AgreeID}$ Stage}
\begin{algorithm}[t]
  \caption{$\texttt{AgreeIDStage}$}
  \label{alg:AgreeIDStage}
  \begin{algorithmic}[1]
    \If{$a_i.\textit{count}=0$}
      \State $a_i.P_p\leftarrow a_i.P_p\cup \{a_j.id \mid a_j\in A_i\wedge a_j.\textit{length}=a_i.\textit{length} \wedge a_j.\textit{stage}=\textit{AgreeID}\}$
    \Else
      \State Execute $\texttt{PCONS}(a_i.S_p)(a_i.\textit{count})$
      \State Execute $\texttt{PCONS}(a_i.P_p)(a_i.\textit{count})$
    \EndIf
    \If{$a_i.\textit{length}>a_i.\textit{elapsed}$}
      \State Execute $\texttt{REL}(a_i.id)(a_i.\textit{elapsed})$
    \Else
      \State $a_i.\textit{elapsed}\leftarrow 0$
      \State $a_i.\textit{count}\leftarrow a_i.\textit{count}+1$
      \If{$\texttt{PCONS}(a_i.S_p)$ and $\texttt{PCONS}(a_i.P_p)$ are finished}
        \State $a_i.S_c\leftarrow$ the output of $\texttt{PCONS}(a_i.S_p)$
        \State $a_i.P_c\leftarrow$ the output of $\texttt{PCONS}(a_i.P_p)$
        \State $a_i.\textit{stage}\leftarrow \textit{MakeGroup}$
      \EndIf
      \State Execute $\texttt{WAIT}()$
    \EndIf
  \end{algorithmic} 
\end{algorithm}

Algorithm \ref{alg:AgreeIDStage} is the pseudo-code of the $\textit{AgreeID}$ stage.
This stage aims to obtain a common ID set among the agents in a group candidate by using consensus algorithm $\texttt{PCONS}$.
The common ID set contains at least $f+1$ agent IDs in the same group candidate.
The agents use the common ID set in order to create a reliable group in the $\textit{MakeGroup}$ stage.
By the behavior of the $\textit{MakeCandidate}$ stage, we have the following observation.

\begin{observation}
\label{obs:AgentsOfSameGroupCandidateHaveSameLengthOfCycle}
Let $a_i$ and $a_j$ be good agents.
If $a_i$ and $a_j$ belong to the same group candidate, for any $\gamma$, they start the first round of $c_i^\gamma$ and $c_j^\gamma$ at the same time.
\end{observation}

In order for agents to efficiently create a reliable group, an agent collects IDs in the same group candidate and makes a consensus on a set of the collected IDs.
In the $\textit{MakeGroup}$ stage, agents create the reliable group by the method using the group candidate as described in Section \ref{sec:Overview}.
If agents make a consensus on $S_p$ to make the common ID set, the output may contain agent IDs not in the same group candidate.
This approach makes the reliable group creation in the $\textit{MakeGroup}$ stage inefficient.
Therefore, in the $\textit{AgreeID}$ stage, agents collect agent IDs in the same group candidate and propose the IDs for consensus.
Since the output contains agent IDs in the same group candidate only, this idea can avoid the above problem.
An agent uses variable $P_p$ to record the IDs collected in the $\textit{AgreeID}$ stage.

In addition, an agent makes a consensus on $S_p$ in this stage.
As we explain in Section \ref{subsubsec:MakeGroupStage}, the output of the consensus is used in the $\textit{MakeGroup}$ stage.

Hereinafter, we explain the detailed behaviors of the $\textit{AgreeID}$ stage.
If an agent $a_i$ executes the first cycle $c_i^\gamma$ of the $\textit{AgreeID}$ stage, that is, $a_i.\textit{count}=0$ holds, $a_i$ collects agent IDs in the same group candidate, say $\textit{GC}$ (Lines 1--2 of Algorithm \ref{alg:AgreeIDStage}).
To be more precise, if $a_i$ witnesses the agent $a_j$ in $\textit{GC}$ at the current node in a round of cycle $c_i^\gamma$, $a_i$ stores $a_j.id$ in $a_i.P_p$.
Agent $a_i$ determines whether $a_j$ belongs to $\textit{GC}$ by checking $a_j.\textit{length}$ and $a_j.\textit{stage}$.
Agent $a_i$ meets all good agents in $\textit{GC}$ and includes their IDs in $a_i.P_p$ by the end of cycle $c_i^\gamma$.

If $a_i$ executes the second or later cycle of the $\textit{AgreeID}$ stage, that is, $a_i.\textit{count}>0$ holds, then $a_i$ makes a consensus on $a_i.S_p$ and $a_i.P_p$ using $\texttt{PCONS}(a_i.S_p)$ and $\texttt{PCONS}(a_i.P_p)$ to obtain the common ID sets with good agents in $\textit{GC}$ (Lines 3--6).
As we mentioned in Section \ref{subsec:MRGC_Idea}, agents simulate one phase of the message-passing model by executing $\texttt{REL}$ during one cycle.
In Algorithm \ref{alg:AgreeIDStage}, when an agent executes $\texttt{PCONS}(S)(q)$ except the last round of a cycle, it shares messages of algorithm $\texttt{PCONS}(S)$ with other agents at the current node. 
If an agent executes $\texttt{PCONS}(S)(q)$ in the last round of a cycle, it computes the $q$-th phase of $\texttt{PCONS}(S)$ using the collected messages.

To simulate $\texttt{PCONS}(a_i.S_p)$ and $\texttt{PCONS}(a_i.P_p)$, $a_i$ meets the other agents in $\textit{GC}$ using $\texttt{REL}$ for a cycle.
If the current round is not the last round of the current cycle, that is, $a_i.\textit{length}>a_i.\textit{elapsed}$ holds, then $a_i$ executes $\texttt{REL}(a_i.id)(a_i.\textit{elapsed})$ to meet the other agents in $\textit{GC}$ in the first cycle, and to execute $\texttt{PCONS}(S)(q)$ with agents in $\textit{GC}$ in and after the second cycle (Lines 7--8).
To make a consensus with good agents in $\textit{GC}$, $a_i$ checks to which group candidate an agent belongs, identically with $P_p$.
Otherwise, that is, if $a_i.\textit{length}=a_i.\textit{elapsed}$ holds, then $a_i$ first checks the status of $\texttt{PCONS}(a_i.S_p)$ and $\texttt{PCONS}(a_i.P_p)$.
If both the consensus instances have finished, $a_i$ stores their outputs in $a_i.S_c$ and $a_i.P_c$, respectively and changes $a_i.\textit{stage}$ into $\textit{MakeGroup}$ (Lines 12--16).
Agent $a_i$ then stays at the current node for one round (Line 17).
If $\textit{GC}$ includes at least $3f+1$ agents, the execution of $\texttt{PCONS}$ by agents in $\textit{GC}$ satisfies the PBC property by Observation \ref{obs:AgentsOfSameGroupCandidateHaveSameLengthOfCycle}.
Therefore, all the good agents in $\textit{GC}$ has identical $P_c$ and $S_c$.
Moreover, since $P_p$ of every good agent contains the IDs of all good agents in $\textit{GC}$, $P_c$ also contains them by PBC Validity 1.

\paragraph{$\textit{MakeGroup}$ Stage}
\label{subsubsec:MakeGroupStage}
\begin{algorithm}[t]
  \caption{$\texttt{MakeGroupStage}$}
  \label{alg:MakeGroupStage}
  \begin{algorithmic}[1]
    \If{$(1/2)\cdot a_i.\textit{length}\geq a_i.\textit{elapsed}$}
      \State Execute $\texttt{REL}(a_i.id)(a_i.\textit{elapsed})$
    \ElsIf{$a_i.\textit{length}>a_i.\textit{elapsed}$}
      \If{$\exists a_j \in A_i[a_j.id=a_i.P_c[a_i.\textit{count} \bmod |a_i.P_c|]]$}
        \State Execute $\texttt{WAIT}()$ 
      \Else
        \State Execute $\texttt{REL}(a_i.id)(a_i.\textit{elapsed})$
      \EndIf
    \Else
      \State $a_i.\textit{elapsed}\leftarrow 0$
      \State $a_i.\textit{count}\leftarrow a_i.\textit{count}+1$
      \State $a_i.D\leftarrow\{a_j.id \mid a_j\in A_i \wedge|a_j.S_c|\geq (8/9)|a_j.S_p| \wedge a_j.\textit{length}=a_i.\textit{length}\wedge a_j.S_c=a_i.S_c\wedge a_j.\textit{stage}=\textit{MakeGroup}\}$
      \If{$|a_i.S_c|\geq (8/9)|a_i.S_p|\wedge|a_i.D|\geq (3/9)|a_i.S_c|$}
        \State $a_i.\textit{gid}\leftarrow \min(a_i.D)$
      \EndIf
      \State Execute $\texttt{WAIT}()$
    \EndIf
  \end{algorithmic} 
\end{algorithm}

Algorithm \ref{alg:MakeGroupStage} is the pseudo-code of the $\textit{MakeGroup}$ stage.
This stage aims to create a reliable group.
An agent $a_i$ has variable $\textit{gid}$ to store its group ID when it becomes a member of a reliable group.
In Algorithm $\texttt{ByzantineGathering}$, which we describe later, since agents do not know $f$, when an agent $a_\ell$ determines whether a reliable group exists at the current node, $a_\ell$ uses $(1/8)|a_\ell.S_p|$ instead of $f$ for its decision.
However, $|a_\ell.S_p|$ differs from the other good agents because every good agent possibly met a different number of Byzantine agents in the $\textit{CollectID}$ stage.
Therefore, in order to be able to recognize a reliable group even if a good agent has any $S_p$, we define a reliable group as follows.

\begin{definition}[Reliable group]
\label{def:ReliableGroup}
A set $\textit{RG}$ of agents is a reliable group if and only if $\textit{RG}$ contains at least $k/8$ good agents and $a_i.\textit{gid}=a_j.\textit{gid}$ holds for any two different good agents $a_i,a_j\in \textit{RG}$.
\end{definition}

Agents behaves differently in the first and second halves of each cycle.
If the current round is in the first half of the current cycle, that is, $(1/2)\cdot a_i.\textit{length}\geq a_i.\textit{elapsed}$ holds, $a_i$ executes $\texttt{REL}(a_i.id)(a_i.\textit{elapsed})$ %
to meet agents in the $\textit{CollectID}$ stage and a reliable group that acts for gathering by Algorithm $\texttt{ByzantineGathering}$ (Lines 1--2 of Algorithm \ref{alg:MakeGroupStage}).

If the current round is in the second half of the current cycle, that is, $a_i.\textit{length}\geq a_i.\textit{elapsed}>(1/2)\cdot a_i.\textit{length}$ holds, $a_i$ creates a reliable group using the common ID set of the $\textit{AgreeID}$ stage (Lines 3--16).
Let $\textit{GC}$ be a group candidate of $a_i$.
As we mentioned in Section \ref{sec:Overview}, agents in $\textit{GC}$ decide a target ID base on the common ID set.
More concretely, they use variables $P_c$ and $\textit{count}$ to decide a target ID.
Since agents in $\textit{GC}$ count the number of cycles from the time they started the $\textit{AgreeID}$ stage, they have the same $\textit{count}$.
Therefore, agents in $\textit{GC}$ decide the same target ID in each cycle of the $\textit{MakeGroup}$ stage.
The strategy for the reliable group creation is as follows.
Agents in $\textit{GC}$ decide a target ID from variables $P_c$ and $\textit{count}$ and searches for the agent with the target ID, say $a_{target}$, using $\texttt{REL}$.
Agent $a_{target}$ stays at the node while the other agents in $\textit{GC}$ search for $a_{target}$.
If agents in $\textit{GC}$ fail to create a reliable group in the last round of the current cycle, they decide on a new target ID from the common ID set using $\textit{count}$ in the first round of the next cycle and repeat the above behavior.
Since an agent updates $\textit{count}$ in the last round of each cycle, these target IDs are different.
Also, since the common ID set contains at least one ID of good agents in $\textit{GC}$ and does not include IDs of good agents not in $\textit{GC}$, good agents choose their IDs as the target ID at least once and create a reliable group by they repeat $f+1$ times.

Agent $a_i$ stores the common ID set in $a_i.P_c$ and achieves the above behavior as follows.
If $a_i$ has the target ID, $a_i$ stays at the node until the last round of the current cycle, otherwise $a_i$ searches for $a_{target}$ using $\texttt{REL}$ until the last round of the current cycle (Lines 3--8).
Then, $a_i$ updates variables for the next cycle in the last round of the cycle (Lines 9--17).

Hereinafter, we explain the detailed behaviors of the second half of the $\textit{MakeGroup}$ stage.
If the current round is not the last round of the current cycle, that is, $a_i.\textit{length}>a_i.\textit{elapsed}$ holds, and either of the following is satisfied, then $a_i$ stays at the current node for one round (Lines 4--5).
\begin{itemize}
  \item[(1)]$a_i$ has the target ID.
  \item[(2)]$a_i$ meets $a_{target}$ at the current node.
\end{itemize}
Agent $a_i$ calculates $a_i.P_c[a_i.\textit{count} \bmod |a_i.P_c|]$ to determine Conditions (1) and (2).
The formula is for $a_i$ to determine the target ID and uses $a_i.P_c$ and $a_i.\textit{count}$ to determine the same target ID among good agents in $\textit{GC}$.
Since agents in $\textit{GC}$ start the $\textit{AgreeID}$ stage at the same time, and they start a cycle at the same time by Observation \ref{obs:LargerXGoodStartTimeAndSmallerXGoodStartTimeSame}, every good agent in $\textit{GC}$ has the same $\textit{count}$.
Furthermore, if $\textit{GC}$ contains at least $3f+1$ agents, every good agent in $\textit{GC}$ has the same $P_c$ by Lemma \ref{lem:AtLeast7/18gGoodCollectlyFinishAgreeID}, which we prove later, and calculates the same target ID.

If the current round is not the last one of the current cycle, that is, $a_i.\textit{length}>a_i.\textit{elapsed}$ holds, and neither Conditions (1) nor (2) is satisfied, $a_i$ executes $\texttt{REL}(a_i.id)(a_i.\textit{elapsed})$ to meet $a_{target}$ (Lines 6--8).

If the current round is the last round of the current cycle, that is, $a_i.\textit{length}=a_i.\textit{elapsed}$ holds, $a_i$ creates a reliable group if possible and stays at the current node for one round (Lines 9--17).
If $a_i$ witnesses an agent $a_j$ that satisfies all of the following, $a_i$ stores $a_j.id$ in variable $D$ to record such IDs (Line 12).
\begin{itemize}
  \item[(a)]$a_j.S_c$ contains at least $(8/9)|a_j.S_p|$ IDs.
  \item[(b)]$a_j$ belongs to $\textit{GC}$.
  \item[(c)]$a_i.S_c$ and $a_j.S_c$ are the same.
  \item[(d)]$a_j$ has started the $\textit{MakeGroup}$ stage.
\end{itemize}
If $a_i.S_c$ contains at least $(8/9)|a_i.S_p|$ IDs and $a_i.D$ contains at least $(3/9)|a_i.S_c|$ IDs, $a_i$ stores the smallest ID in $a_i.D$ in $a_i.\textit{gid}$ as the group ID (Lines 13--15).

By the above behavior, assuming that $a_i$ stores the group ID in $a_i.\textit{gid}$ in the last round $r$ of some cycle, we can guarantee that the agents in $\textit{GC}$ create the reliable group.
In round $r$, there exist at least $(3/9)|a_i.S_c|$ agents that satisfy Conditions (a)--(d) at the current node, and $(3/9)|a_i.S_c|\geq (3/9)(8/9)|a_i.S_p|=(8/27)|a_i.S_p|\geq (8/27)g$ holds by Line 13.
If $\textit{GC}$ contains at least $3f+1$ agents, every good agent in $\textit{GC}$ has the same $S_c$.
Furthermore, agents at the current node have the same $D$.
Therefore, at least $(8/27)g-f$ good agents reach the same decision.
By $k\geq 9f+8$ and $g\geq 8f+8$, $(8/27)g-f=(27/216)g+(37/216)g-f\geq (27/216)g+(37/216)(8f+8)-f=(27/216)g+(80/216)f+296/216>(27/216)(g+f)=k/8$ holds.
Hence, in round $r$, at least $k/8$ good agents, including $a_i$, store the same group ID and create a reliable group.

\subsubsection{Correctness and Complexity Analysis}
In this subsection, we prove the correctness and complexity of $\texttt{MakeReliableGroup}$.
First, we prove that, when agent $a_i$ executes $\texttt{REL}(a_i.id)$ throughout a cycle, it can meet all agents.
Note that, at the beginning of a cycle, $a_i$ determines whether it executes $\texttt{REL}(a_i.id)$ or waits for the whole cycle.
If $a_i$ starts $\texttt{REL}(a_i.id)$, it stops the procedure at the middle of a cycle only if $a_i$ satisfies Line 4 of Algorithm \ref{alg:MakeGroupStage}.
Even in this case, $a_i$ executes $\texttt{REL}(a_i.id)$ for at least the half of the cycle.
We say ``$a_i$ executes $\texttt{REL}(a_i.id)$ without interruption'' if $a_i$ executes $\texttt{REL}(a_i.id)$ throughout a cycle.

\begin{lemma}
\label{lem:GoodMeetAllGood}
Let $a_i$ be a good agent and $c_i^\gamma$ be a cycle in which $a_i$ starts $\texttt{REL}(a_i.id)$ at the beginning of the first round of this cycle.
If $a_i$ executes $\texttt{REL}(a_i.id)$ until the last round of cycle $c_i^\gamma$ without interruption, then $a_i$ meets all good agents during cycle $c_i^\gamma$.
\end{lemma}
\begin{proof}
By the behavior of $\texttt{MakeReliableGroup}$, since $a_i$ starts $\texttt{REL}(a_i.id)$ in cycle $c_i^\gamma$, $a_i$ satisfies Line 8 of Algorithm \ref{alg:CollectIDStage} no later than the first round of cycle $c_i^\gamma$ and thus $|c_i^\gamma|\geq 2\cdot (\TIME{REL}(a_i.id)+1)$ holds.
Let round $r$ be the first round of cycle $c_i^\gamma$.
Let $a_j$ be a good agent other than $a_i$ and $c_j^\varepsilon$ be a cycle of $a_j$ that includes round $r$.
We prove that $a_i$ and $a_j$ meet during cycle $c_i^\gamma$.
We consider two cases, $|c_i^\gamma|\geq |c_j^\varepsilon|$ and $|c_i^\gamma|<|c_j^\varepsilon|$.

First, we consider the case $|c_i^\gamma|\geq |c_j^\varepsilon|$.
In this case, by the behavior of $\texttt{MakeReliableGroup}$, $a_j$ starts the $\textit{AgreeID}$ stage no later than $a_i$.
Therefore, by Observation \ref{obs:LargerXGoodStartTimeAndSmallerXGoodStartTimeSame}, $a_i$ and $a_j$ start cycles $c_i^\gamma$ and $c_j^\varepsilon$ at the same time at the beginning of round $r$.
Furthermore, when $a_i$ starts cycle $c_i^{\gamma+1}$, $a_j$ also starts $c_j^{\varepsilon+\alpha+1}$ at the same time for a non-negative integer $\alpha$.
In other words, $a_j$ finishes all cycles $c_j^\varepsilon,\dots,c_j^{\varepsilon+\alpha}$ by the time $a_i$ finishes cycle $c_i^\gamma$.
By the behavior of $\texttt{MakeReliableGroup}$, in each cycle, $a_j$ executes $\texttt{REL}(a_j.id)$ for at least the half of the cycle or stays at the current node.
We consider two cases.
First, we consider that $a_j$ stays at the current node in every cycle $c_j^\varepsilon,\dots,c_j^{\varepsilon+\alpha}$.
In this case, $a_j$ stays at the current node throughout all cycles $c_j^\varepsilon,\dots,c_j^{\varepsilon+\alpha}$, that is, cycle $c_i^\gamma$.
Since $|c_i^\gamma|\geq 2\cdot (\TIME{REL}(a_i.id)+1)$ holds, and $a_i$ visits all nodes in $\TIME{REL}(a_i.id)$ rounds after the first round of cycle $c_i^\gamma$ by Lemma \ref{lem:RendezvousAlgorithm}, $a_i$ and $a_j$ meet by the end of cycle $c_i^\gamma$.
Next, we consider that $a_j$ executes $\texttt{REL}(a_j.id)$ in some cycle $c_j^{\varepsilon+\beta}$ ($0\leq\beta\leq\alpha$).
In this case, $a_j$ executes $\texttt{REL}(a_j.id)$ for at least $(1/2)\cdot |c_j^{\varepsilon+\beta}|$ rounds from the first round of cycle $c_j^{\varepsilon+\beta}$.
Since $a_j$ satisfies Line 8 of Algorithm \ref{alg:CollectIDStage} no later than the first round of cycle $c_j^{\varepsilon+\beta}$, $|c_j^{\varepsilon+\beta}|\geq 2\cdot (\TIME{REL}(a_j.id)+1)$ holds.
Thus, $a_i$ and $a_j$ execute $\texttt{REL}(a_i.id)$ and $\texttt{REL}(a_j.id)$ at the same time for at least $\TIME{REL}(\min(a_i.id,a_j.id))$ rounds.
Therefore, $a_i$ and $a_j$ meet by the end of cycle $c_i^\gamma$ by Lemma \ref{lem:RendezvousAlgorithm}.

Next, we consider the case $|c_i^\gamma|<|c_j^\varepsilon|$.
By Observation \ref{obs:LargerXGoodStartTimeAndSmallerXGoodStartTimeSame}, when $a_j$ starts cycle $c_j^{\varepsilon+1}$, $a_i$ starts cycle $c_i^{\gamma'}$ ($\gamma'>\gamma$).
Thus, the period of cycle $c_i^\gamma$ is completely included in the period of cycle $c_j^\varepsilon$.
By Lemma \ref{lem:RendezvousAlgorithm}, $a_i$ visits all nodes in $\TIME{REL}(a_i.id)$ rounds after the first round of cycle $c_i^\gamma$.
Therefore, since $|c_i^\gamma|\geq 2\cdot (\TIME{REL}(a_i.id)+1)$ holds, when $a_j$ stays at the current round throughout the first half of cycle $c_i^\gamma$, $a_i$ and $a_j$ meet by the end of cycle $c_i^\gamma$.
Also, when $a_j$ executes $\texttt{REL}(a_j.id)$ throughout the first half of cycle $c_i^\gamma$, $a_i$ and $a_j$ execute $\texttt{REL}(a_i.id)$ and $\texttt{REL}(a_j.id)$ at the same time for at least $\TIME{REL}(\min(a_i.id,a_j.id))$ rounds.
Thus, $a_i$ and $a_j$ meet by the end of cycle $c_i^\gamma$ by Lemma \ref{lem:RendezvousAlgorithm}.
On the other hand, when $a_j$ is in the $\textit{MakeGroup}$ stage, it may interrupt $\texttt{REL}(a_j.id)$ in the fist half of cycle $c_i^\gamma$.
In this case, $a_j$ stays at the current node throughout the last half of cycle $c_i^\gamma$.
During the last half of cycle $c_i^\gamma$, $a_i$ keeps executing $\texttt{REL}(a_i.id)$.
Thus, from the above discussion, $a_i$ and $a_j$ meet by the end of cycle $c_i^\gamma$.
Hence, the lemma holds.
\end{proof}

In the $\textit{CollectID}$ stage, an agent $a_i$ executes $\texttt{REL}(a_i.id)$ without interruption during the cycle whose length is at least $2\cdot (\TIME{REL}(a_i.id)+1)$.
Since $a_i$ meets all good agents by the end of the cycle, we have the following corollary.

\begin{corollary}
\label{cor:GoodKnowAllGoodIds}
For good agent $a_i$, if $a_i.\textit{stage}\in \{\textit{MakeCandidate}$, $\textit{AgreeID}$, $\textit{MakeGroup}\}$ holds, $a_i.S_p$ contains IDs of all good agents and hence $k=g+f\geq |a_i.S_p|\geq g$ holds.
\end{corollary}

By $k\geq 9f+8$, $g\geq 8f+8$, and Corollary \ref{cor:GoodKnowAllGoodIds}, we have the following corollary.

\begin{corollary}
\label{cor:g_geq_(8/9)k_geq_(8/9)Sp}
For good agent $a_i$, if $a_i.\textit{stage}\in \{\textit{MakeCandidate}$, $\textit{AgreeID}$, $\textit{MakeGroup}\}$ holds, $g>(8/9)k\geq (8/9)|a_i.S_p|$ holds.
\end{corollary}

Next, we consider the $\textit{MakeCandidate}$ stage.
Let $a_{max}$ be the good agent with the largest ID.
Regarding this stage, we clarify the following two facts:
(1) All good agents finish the $\textit{MakeCandidate}$ stage in some bounded rounds, and
(2) At least $(7/18)$g good agents start the $\textit{AgreeID}$ stage at the same time.
We prove (1) with Lemma \ref{lem:LastTransitionAgentIsAgentMax} to Lemma \ref{lem:AllGoodFinishMCInOt_REL}, and (2) with Lemma \ref{lem:AgentsSendMessageBySatisfyingAlgorithmMakeCandidate4} to Corollary \ref{cor:7/18gGoodStartAgreeIDTogether}.

\begin{lemma}
\label{lem:LastTransitionAgentIsAgentMax}
Let $c_{max}^\gamma$ be the first cycle of the $\textit{MakeCandidate}$ stage of $a_{max}$.
Every good agent $a_i$ executes $a_i.\textit{stage}\leftarrow \textit{AgreeID}$ by the end of cycle $c_{max}^{\gamma+1}$.
\end{lemma}
\begin{proof}
First, we prove that every good agent $a_i$ executes $a_i.\textit{ready}\leftarrow\textit{True}$ by the end of the first round of cycle $c_{max}^\gamma$.
Since $a_{max}$ has the largest ID among good agents, $a_{max}$ starts the $\textit{MakeCandidate}$ stage latest among good agents.
Therefore, all good agents start the $\textit{MakeCandidate}$ stage by the time $a_{max}$ starts that.
We consider two cases.
First, we consider the case that $a_i$ is in the $\textit{MakeCandidate}$ stage at the beginning of cycle $c_{max}^\gamma$.
By Observation \ref{obs:XsOfCollectIDandPreparationAreSame}, $a_i.\textit{length}=a_{max}.\textit{length}$ holds in cycle $c_{max}^\gamma$ and $a_i$ starts its cycle at the same time as $a_{max}$.
Since $a_{max}$ satisfies Line 8 of Algorithm \ref{alg:CollectIDStage} before cycle $c_{max}^\gamma$, $a_{max}.\textit{length}\geq 4\cdot(\TIME{REL}(a_{max}.id)+1)$ holds at the beginning of cycle $c_{max}^\gamma$.
Therefore, $a_i.\textit{length}\geq 4\cdot(\TIME{REL}(a_{max}.id)+1)$ holds at the beginning of cycle $c_{max}^\gamma$.
Since $a_i.S_p$ contains IDs of all good agents by Corollary \ref{cor:GoodKnowAllGoodIds}, it contains at least $g$ IDs no larger than $a_{max}.id$.
Since $g>(8/9)|a_i.S_p|$ holds by Corollary \ref{cor:g_geq_(8/9)k_geq_(8/9)Sp}, $a_i$ satisfies Line 3 of Algorithm \ref{alg:MakeCandidateStage} at the beginning of cycle $c_{max}^\gamma$.
Thus, $a_i$ executes $a_i.\textit{ready}\leftarrow\textit{True}$ by the end of the first round of cycle $c_{max}^\gamma$.
Next, we consider the case that $a_i$ is in either the $\textit{AgreeID}$ stage or the $\textit{MakeGroup}$ stage at the beginning of cycle $c_{max}^\gamma$.
Since $a_i$ has finished the $\textit{MakeCandidate}$ stage before cycle $c_{max}^\gamma$, $a_i$ satisfies Line 8 of Algorithm \ref{alg:MakeCandidateStage}, that is, $|a_i.R|\geq (6/9)|a_i.S_p|$ holds.
Therefore, $a_i$ satisfies Line 4 of Algorithm \ref{alg:MakeCandidateStage} and thus $a_i$ executes $a_i.\textit{ready}\leftarrow \textit{True}$ before cycle $c_{max}^\gamma$.

Next, we prove that every good agent $a_i$ executes $a_i.\textit{stage}\leftarrow \textit{AgreeID}$ by the end of cycle $c_{max}^{\gamma+1}$.
Consider the situation that good agent $a_i$ has not yet executed $a_i.\textit{stage}\leftarrow \textit{AgreeID}$ at the beginning of cycle $c_{max}^{\gamma+1}$.
Let $a_j$ be a good agent.
First, we consider the case that $a_j.\textit{ready}=\textit{True}$ holds at the beginning of cycle $c_{max}^\gamma$.
By Lemma \ref{lem:GoodMeetAllGood}, since $a_i$ executes $\texttt{REL}(a_i.id)$ throughout cycle $c_{max}^\gamma$, $a_i$ meets $a_j$ by the last round of cycle $c_{max}^\gamma$.
Thus, $a_i.R$ contains $a_j.id$ before cycle $c_{max}^{\gamma+1}$.
Next, we consider the case that $a_j.\textit{ready}=\textit{False}$ holds at the beginning of cycle $c_{max}^\gamma$.
In this case, $a_j$ executes the $\textit{MakeCandidate}$ stage during cycle $c_{max}^\gamma$ and executes $a_j.\textit{ready}\leftarrow\textit{True}$ by the end of the first round of the cycle.
By Observation \ref{obs:XsOfCollectIDandPreparationAreSame}, $a_i$ and $a_j$ start their cycles at the same time and have the same length of their cycles.
Since $|c_{max}^\gamma|\geq 4\cdot (\TIME{REL}(a_{max}.id)+1)$ holds, $a_i$ and $a_j$ execute $\texttt{REL}(a_i.id)$ and $\texttt{REL}(a_j.id)$ at the same time for at least $\TIME{REL}(\min(a_i.id,a_j.id))$ rounds after the first round of cycle $c_{max}^\gamma$.
Therefore, by Lemma \ref{lem:RendezvousAlgorithm}, $a_i$ meets $a_j$ at least once after $a_j$ executes $a_j.\textit{ready}\leftarrow\textit{True}$.
This implies that $a_i.R$ contains $a_j.id$ before cycle $c_{max}^{\gamma+1}$.
From these two cases, $a_i.R$ contains at least $g$ IDs before cycle $c_{max}^{\gamma+1}$.
Since $g>(8/9)|a_i.S_p|$ holds by Corollary \ref{cor:g_geq_(8/9)k_geq_(8/9)Sp}, $a_i$ satisfies Line 8 of Algorithm \ref{alg:MakeCandidateStage} at the beginning of cycle $c_{max}^{\gamma+1}$.
Since $a_{max}$ doubles $a_{max}.\textit{length}$ in the last round of cycle $c_{max}^\gamma$, $|c_{max}^{\gamma+1}|$ is identical to the length of a cycle of a good agent that executes the $\textit{MakeCandidate}$ stage in cycle $c_{max}^{\gamma+1}$.
Therefore, $a_i$ executes $a_i.\textit{stage}\leftarrow \textit{AgreeID}$ in the last round of cycle $c_{max}^{\gamma+1}$.

Hence, the lemma holds.
\end{proof}

In the following lemma, we calculate the maximum value of $\textit{length}$ of a good agent.

\begin{lemma}
\label{lem:MaxLength}
For any good agent $a_i$, $a_i.\textit{length}$ is less than $32\cdot (\TIME{REL}(a_{max}.id)+1)$.
\end{lemma}
\begin{proof}
Let $c_{max}^\gamma$ be the first cycle of the $\textit{MakeCandidate}$ stage of $a_{max}$.
Each good agent $a_i$ updates $a_i.\textit{length}$ in the last round of a cycle of the $\textit{CollectID}$ and $\textit{MakeCandidate}$ stages, but not in the $\textit{AgreeID}$ and $\textit{MakeGroup}$ stages. 
Also, by Lemma \ref{lem:LastTransitionAgentIsAgentMax}, $a_i$ executes $a_i.\textit{stage}\leftarrow\textit{AgreeID}$ by the end of cycle $c_{max}^{\gamma+1}$.
Thus, to calculate the maximum value of $a_i.\textit{length}$, it is enough to calculate the value of $a_{last}.\textit{length}$ at the beginning of cycle $c_{max}^{\gamma+2}$ for a good agent $a_{last}$ that starts the $\textit{AgreeID}$ stage latest.
Let $\textit{length}_{last}^\alpha$ be the value of $a_{last}.\textit{length}$ at the beginning of cycle $c_{max}^\alpha$ for a positive integer $\alpha$.
When $a_{last}$ starts the $\textit{MakeCandidate}$ stage in cycle $c_{max}^\gamma$, $a_{last}$ satisfies Line 8 of Algorithm \ref{alg:CollectIDStage} (i.e., $2\cdot (\TIME{REL}(a_{max}.id)+1)\leq \textit{length}_{last}^{\gamma-1}$ holds at the beginning of cycle $c_{max}^{\gamma-1}$).
On the other hand, since $a_{last}$ does not satisfy Line 8 of Algorithm \ref{alg:CollectIDStage} at the beginning of cycle $c_{max}^{\gamma-2}$, $2\cdot (\TIME{REL}(a_{max}.id)+1)>\textit{length}_{last}^{\gamma-2}$ holds.
Hence, $2^4\cdot 2\cdot (\TIME{REL}(a_{max}.id)+1)=32\cdot (\TIME{REL}(a_{max}.id)+1)>\textit{length}_{last}^{\gamma+2}$ holds at the beginning of cycle $c_{max}^{\gamma+2}$.
\end{proof}

\begin{lemma}
\label{lem:AllGoodFinishMCInOt_REL}
All good agents finish the $\textit{CollectID}$ stage in $O(\TIME{REL}(a_{max}.id))$ rounds after starting $\texttt{MakeReliableGroup}$.
Furthermore, all good agents finish the $\textit{MakeCandidate}$ stage in $O(\TIME{REL}(a_{max}.id))$ rounds after starting $\texttt{MakeReliableGroup}$.
\end{lemma}
\begin{proof}
Let $c_{max}^\gamma$ be the first cycle of the $\textit{MakeCandidate}$ stage of $a_{max}$.
By Lemma \ref{lem:LastTransitionAgentIsAgentMax}, all good agents execute $\textit{stage}\leftarrow\textit{AgreeID}$ by the end of cycle $c_{max}^{\gamma+1}$.
Let $a_{last}$ be a good agent that finishes the $\textit{MakeCandate}$ stage latest.
To prove this lemma, it is enough to analyze the time $t_{mc}$ that are required for $a_{last}$ to finish the $\textit{MakeCandidate}$ stage.
Agent $a_{last}$ repeats a cycle and its length is $a_{last}.\textit{length}$.
Furthermore, $a_{last}$ doubles $a_{last}.\textit{length}$ in the last round of a cycle of the $\textit{CollectID}$ and $\textit{MakeCandidate}$ stages.
By Lemma \ref{lem:MaxLength}, $|c_{max}^{\gamma+2}|<32\cdot (\TIME{REL}(a_{max}.id)+1)$ holds.
Thus, since $|c_{max}^\alpha|$ is represented as $2^{\alpha-1}\cdot T_{ini}$ for a positive integer $\alpha$, $t_{mc}$ is at most $T_{ini}+2\cdot T_{ini}+4\cdot T_{ini}+\dots+2^{\gamma+1}\cdot T_{ini}=(2^{\gamma+2}-1)\cdot T_{ini}=|c_{max}^{\gamma+2}|-T_{ini}<32\cdot (\TIME{REL}(a_{max}.id)+1)-T_{ini}$.
Therefore, $a_{last}$ finishes the $\textit{MakeCandidate}$ stage in $O(\TIME{REL}(a_{max}.id))$ rounds after starting $\texttt{MakeReliableGroup}$.
Hence, all good agents finish the $\textit{MakeCandidate}$ stage in $O(\TIME{REL}(a_{max}.id))$ rounds after starting $\texttt{MakeReliableGroup}$, which implies that they also finish the $\textit{CollectID}$ stage in $O(\TIME{REL}(a_{max}.id))$ rounds.
\end{proof}

From now on, we will prove that at least $(7/18)g$ good agents start the $\textit{AgreeID}$ stage at the same time.
First, we focus on the situation where the first good agent becomes ready to transition into the $\textit{AgreeID}$ stage.

\begin{lemma}
\label{lem:AgentsSendMessageBySatisfyingAlgorithmMakeCandidate4}
Let $a_i$ be the first good agent that executes $\textit{ready}\leftarrow \textit{True}$.
Agent $a_i$ executes $a_i.\textit{ready}\leftarrow\textit{True}$ by satisfying Line 3 of Algorithm \ref{alg:MakeCandidateStage}.
\end{lemma}
\begin{proof}
We prove this lemma by contradiction.
Let $c_i^\gamma$ be the cycle in which $a_i$ executes $a_i.\textit{ready}\leftarrow\textit{True}$.
At the beginning of cycle $c_i^\gamma$, $|a_i.R|\leq f$ holds because only Byzantine agents can execute $\textit{ready}\leftarrow\textit{True}$ before cycle $c_i^\gamma$.
On the other hand, $a_i$ should satisfy $|a_i.R|\geq (4/9)|a_i.S_p|$ at the beginning of cycle $c_i^\gamma$ to execute $a_i.\textit{ready}=\textit{True}$ without satisfying Line 3 of Algorithm \ref{alg:MakeCandidateStage}.
Since $k\geq |a_i.S_p|\geq g$ holds by Corollary \ref{cor:GoodKnowAllGoodIds}, $(4/9)|a_i.S_p|\geq (4/9)g\geq (4/9)(8f+8)=(32/9)f+32/9>f$ holds by $g\geq 8f+8$.
This is a contradiction.
\end{proof}

By Lemma \ref{lem:AgentsSendMessageBySatisfyingAlgorithmMakeCandidate4}, at least one good agent $a_i$ executes $a_i.\textit{ready}\leftarrow \textit{True}$ by satisfying Line 3 of Algorithm \ref{alg:MakeCandidateStage}.
In the following lemma, we check $S_p$ of good agents that execute the $\textit{MakeCandidate}$ stage at the beginning of the cycle when $a_i$ executes $a_i.\textit{ready}\leftarrow \textit{True}$.

\begin{lemma}
\label{lem:WhenFirstBroadcastGoodGoodHave7/9Sp}
Let $a_i$ be the first good agent that executes $\textit{ready}\leftarrow\textit{True}$ and $c_i^\gamma$ be the cycle in which $a_i$ executes that.
Let $a_j$ be a good agent that is in the $\textit{MakeCandidate}$ stage in cycle $c_i^\gamma$.
At the beginning of cycle $c_i^\gamma$, $a_j.S_p$ contains at least $(7/9) |a_j.S_p|$ IDs of good agents that started the $\textit{MakeCandidate}$ stage by the beginning of cycle $c_i^\gamma$.
\end{lemma}
\begin{proof}
By Corollary \ref{cor:GoodKnowAllGoodIds}, $a_i.S_p$ and $a_j.S_p$ contain IDs of all good agents at the beginning of cycle $c_i^\gamma$.
Therefore, let $f_i$ (resp. $f_j$) be the number of IDs of Byzantine agents in $a_i.S_p$ (resp. $a_j.S_p$), and then $|a_i.S_p|=g+f_i$ (resp. $|a_j.S_p|=g+f_j$) holds.
By Lemma \ref{lem:AgentsSendMessageBySatisfyingAlgorithmMakeCandidate4}, at the beginning of cycle $c_i^\gamma$, $a_i.S_p$ contains at least $(8/9)|a_i.S_p|$ IDs of agents that started the $\textit{MakeCandidate}$ stage by the beginning of cycle $c_i^\gamma$.
Thus, $a_i.S_p$ contains at least $(8/9)|a_i.S_p|-f_i$ IDs of good agents.
By $f\geq f_i$, $k\geq 9f+8$, and $g\geq 8f+8$, $(8/9)|a_i.S_p|-f_i=(8/9)(g+f_i)-f_i=(1/9)(8g+8f_i-9f_i)=(1/9)(8g-f_i)\geq (1/9)(8g-f)=(1/9)(7g+g-f)\geq (1/9)(7g+8f+8-f)=(1/9)(7g+7f+8)>(7/9)(g+f)$ holds.
Therefore, by $f\geq f_j$, at the beginning of cycle $c_i^\gamma$, $a_j.S_p$ contains at least $(7/9)(g+f)\geq (7/9)(g+f_j)=(7/9)|a_j.S_p|$ IDs of good agents that started the $\textit{MakeCandidate}$ stage by the beginning of cycle $c_i^\gamma$.
\end{proof}

In the following lemma, we check $R$ of good agents that execute the $\textit{MakeCandidate}$ stage when a good agent $a_i$ executes $a_i.\textit{endMakeCandidate}\leftarrow\textit{True}$.

\begin{lemma}
\label{lem:IfReceived6/9SpMessagesGood4/9SpMessages}
Let $a_i$ be a good agent that executes $\textit{endMakeCandidate}\leftarrow\textit{True}$, and $c_i^\gamma$ be the cycle in which $a_i$ executes that.
Let $a_j$ be a good agent that is in the $\textit{MakeCandidate}$ stage in cycle $c_i^\gamma$.
At the beginning of cycle $c_i^\gamma$, $a_j.R$ contains at least $(4/9)|a_j.S_p|$ IDs of good agents.
\end{lemma}
\begin{proof}
At the beginning of cycle $c_i^\gamma$, $a_i.R$ contains at least $(6/9)|a_i.S_p|$ IDs of agents.
Thus, $a_i.R$ contains at least $(6/9)|a_i.S_p|-f$ IDs of good agents.
Since $k\geq |a_i.S_p|\geq g$ holds by Corollary \ref{cor:GoodKnowAllGoodIds}, $(6/9)|a_i.S_p|-f\geq (6/9)g-f$ holds.
By $g\geq 8f+8$, $(6/9)g-f=(1/9)(4g+2g-9f)>(1/9)(4g+16f-9f)=(1/9)(4g+7f)>(4/9)(g+f)$ holds.
Since $k\geq |a_j.S_p|\geq g$ holds, $a_j.R$ contains at least $(4/9)(g+f)\geq (4/9)|a_j.S_p|$ IDs of agents at the beginning of cycle $c_i^\gamma$.
\end{proof}

By Lemma \ref{lem:LastTransitionAgentIsAgentMax}, every good agent starts the $\textit{AgreeID}$ stage by the time $a_{max}$ starts that.
In the following lemma, we show that several good agents start the $\textit{AgreeID}$ stage at the same time.

\begin{lemma}
\label{lem:AtLeast7/9gGoodTransitToAgreeIDInEitherOfTheTwoRounds}
Let $a_{ini}$ be the first good agent that starts the $\textit{AgreeID}$ stage and $c_{ini}^\gamma$ be a cycle in which $a_{ini}$ starts the $\textit{AgreeID}$ stage.
At least $(7/9)g$ good agents start the $\textit{AgreeID}$ stage in the first round of cycle $c_{ini}^\gamma$ or cycle $c_{ini}^{\gamma+1}$.
\end{lemma}
\begin{proof}
First, we prove that at least $(7/9)g$ good agents have started the $\textit{MakeCandidate}$ stage before cycle $c_{ini}^{\gamma-2}$.
By the assumption of $a_{ini}$, every good agent executes the $\textit{CollectID}$ stage or the $\textit{MakeCandidate}$ stage at the beginning of cycle $c_{ini}^{\gamma-1}$.
Thus, by Observation \ref{obs:XsOfCollectIDandPreparationAreSame}, every good agent has the same $\textit{length}$ and starts its cycle at the same time at the beginning of cycle $c_{ini}^{\gamma-1}$.
Since $a_{ini}$ starts the $\textit{AgreeID}$ stage at the beginning of cycle $c_{ini}^\gamma$, $a_{ini}$ satisfies Line 8 of Algorithm \ref{alg:MakeCandidateStage} at the beginning of cycle $c_{ini}^{\gamma-1}$.
That is, $|a_{ini}.R|\geq (6/9)|a_{ini}.S_p|$ holds.
Since $k\geq |a_{ini}.S_p|\geq g$ holds by Corollary \ref{cor:GoodKnowAllGoodIds}, $|a_{ini}.R|\geq (6/9)|a_{ini}.S_p|\geq (6/9)g\geq (6/9)(8f+8)$ holds by $g\geq 8f+8$.
Since $f$ Byzantine agents exist in the network, at least $(6/9)(8f+8)-f>1$ good agents execute $\textit{ready}\leftarrow\textit{True}$ in cycle $c_{ini}^{\gamma-1}$ or earlier.
This implies, by Lemma \ref{lem:WhenFirstBroadcastGoodGoodHave7/9Sp}, that $a_{ini}.S_p$ contains at least $(7/9)|a_{ini}.S_p|$ IDs of good agents that have started the $\textit{MakeCandidate}$ stage before cycle $c_{ini}^{\gamma-2}$.
That is, at least $(7/9)|a_{ini}.S_p|\geq (7/9)g$ good agents have started the $\textit{MakeCandidate}$ stage before cycle $c_{ini}^{\gamma-2}$. 

Let $A_{em}$ be a set of good agents that are in the $\textit{MakeCandidate}$ stage in the first round of cycle $c_{ini}^{\gamma-1}$.
From the above discussion, $|A_{em}|\geq (7/9)g$ holds.
The set $A_{em}$ are divided into the following two sets $A_1$ and $A_2$:
$A_1$ is a set of agents that satisfies Line 8 of Algorithm \ref{alg:MakeCandidateStage} in the first round of cycle $c_{ini}^{\gamma-1}$, and $A_2=A_{em}\setminus A_1$.
Note that agents in $A_1$ start the $\textit{AgreeID}$ stage in the first round of cycle $c_{ini}^\gamma$.
Consider an arbitrary agent $a_{em}$ in $A_{em}$.
Since $a_{ini}$ executes $a_{ini}.\textit{endMakeCandidate}\leftarrow\textit{True}$ in the first round of cycle $c_{ini}^{\gamma-1}$, $|a_{em}.R|$ contains at least $(4/9)|a_{em}.S_p|$ IDs at the beginning of cycle $c_{ini}^{\gamma-1}$ by Lemma \ref{lem:IfReceived6/9SpMessagesGood4/9SpMessages}.
Therefore, $a_{em}$ satisfies Line 4 of Algorithm \ref{alg:MakeCandidateStage} and executes $a_{em}.\textit{ready}\leftarrow\textit{True}$ in the first round of cycle $c_{ini}^{\gamma-1}$ if $a_{em}.\textit{ready}=\textit{False}$ holds.
That is, $a_{em}.\textit{ready}=\textit{True}$ holds in the first round of cycle $c_{ini}^{\gamma-1}$.
Consider an agent $a_i$ in $A_2$.
Since $a_i$ meets all good agents by the end of cycle $c_{ini}^{\gamma-1}$ by Lemma \ref{lem:GoodMeetAllGood}, $|a_i.R|$ contains at least $|A_{em}|\geq (7/9)g$ IDs at the beginning of cycle $c_{ini}^\gamma$.
Since $g>(8/9)k$ by Corollary \ref{cor:g_geq_(8/9)k_geq_(8/9)Sp}, $(7/9)g>(7/9)\cdot (8/9)k=(56/81)k>(6/9)k$ holds.
Thus, $a_i$ satisfies Line 8 of Algorithm \ref{alg:MakeCandidateStage} in the first round of cycle $c_{ini}^\gamma$ and executes $a_i.\textit{stage}\leftarrow\textit{AgreeID}$ in the last round of cycle $c_{ini}^\gamma$.
Therefore, agents in $A_1$ start the $\textit{AgreeID}$ stage at the beginning of $c_{ini}^\gamma$ and agents in $A_2$ start that at the beginning of $c_{ini}^{\gamma+1}$.
Hence, at least $(7/9)g$ good agents start the $\textit{AgreeID}$ stage at the beginning of cycle $c_{ini}^\gamma$ or cycle $c_{ini}^{\gamma+1}$.
\end{proof}

By Lemma \ref{lem:AtLeast7/9gGoodTransitToAgreeIDInEitherOfTheTwoRounds}, we have the following corollary.

\begin{corollary}
\label{cor:7/18gGoodStartAgreeIDTogether}
At least $(7/18)g$ good agents start the $\textit{AgreeID}$ stage at the same time.
\end{corollary}

Next, we consider the $\textit{AgreeID}$ stage.
Regarding this stage, we check $P_p$ of good agents in the $\textit{AgreeID}$ stage and the execution of $\texttt{PCONS}$ for both $S_p$ and $P_p$.

\begin{lemma}
\label{lem:AtLeast7/18gGoodHaveAtLeast7/18gIDsInPp}
Let $\textit{GC}$ be a group candidate, $a_i$ be a good agent in $\textit{GC}$, and $c_i^\gamma$ be a cycle in which $a_i$ starts the $\textit{AgreeID}$ stage.
Variable $a_i.P_p$ contains at least IDs of all good agents in $\textit{GC}$ at the beginning of cycle $c_i^{\gamma+1}$.
\end{lemma}
\begin{proof}
Since $a_i.\textit{count}=0$ holds at the beginning of cycle $c_i^\gamma$ by the behavior of $\texttt{MakeReliableGroup}$, $a_i$ satisfies Line 1 of Algorithm \ref{alg:AgreeIDStage} at the beginning of cycle $c_i^\gamma$ and thus starts collecting IDs of agents in $\textit{GC}$ using $\texttt{REL}(a_i.id)$.
All good agents in $\textit{GC}$ have the same length of a cycle by Observation \ref{obs:CountsOfConsensusAndGathringAreSame}.
Also, since all good agents in $\textit{GC}$ start the $\textit{AgreeID}$ stage at the same time by Definition \ref{def:GroupCandidate}, they execute $\textit{stage}\leftarrow\textit{AgreeID}$ before cycle $c_i^\gamma$.
Therefore, since $a_i$ meets all good agents by the end of cycle $c_i^\gamma$ by Lemma \ref{lem:GoodMeetAllGood}, $a_i.P_p$ contains at least IDs of all good agents in $\textit{GC}$ at the beginning of cycle $c_i^{\gamma+1}$.
\end{proof}

\begin{lemma}
\label{lem:AtLeast7/18gGoodCollectlyFinishAgreeID}
Let $\textit{GC}$ be a group candidate, $a_i$ be a good agent in $\textit{GC}$, and $c_i^\gamma$ be a cycle in which $a_i$ starts the $\textit{AgreeID}$ stage.
If at least $(7/18)g$ good agents belong to $\textit{GC}$, all good agents in $\textit{GC}$ start the $\textit{MakeGroup}$ stage in $O(f)$ cycles after cycle $c_i^\gamma$.
Furthermore, the executions of both $\texttt{PCONS}(a_i.S_p)$ and $\texttt{PCONS}(a_i.P_p)$ satisfy the PBC property.
\end{lemma}
\begin{proof}
First, we show that all good agents in $\textit{GC}$ can simulate $\texttt{PCONS}$.
Algorithm $\texttt{MakeReliableGroup}$ tries to realize one phase in the Byzantine synchronous message-passing model by one cycle.
By Observations \ref{obs:CountsOfConsensusAndGathringAreSame} and \ref{obs:AgentsOfSameGroupCandidateHaveSameLengthOfCycle}, all good agents in $\textit{GC}$ have the same length of a cycle and start a cycle at the same time.
By the behavior of $\texttt{MakeReliableGroup}$, $a_i$ executes $\texttt{REL}(a_i.id)$ every cycle of the $\textit{AgreeID}$ stage.
Also, by Lemma \ref{lem:GoodMeetAllGood}, $a_i$ meets all good agents in $\textit{GC}$ by the end of the cycle.
Thus, for a message $\textit{msg}$ that a good agent $s$ in $\textit{GC}$ send in a phase, its destination agent $d$ receives $\textit{msg}$ from $s$ by hand when they meet during the phase and $d$ knows the ID of $s$ at that time.
This behavior follows the definition of the synchronous message-passing model that $\texttt{PCONS}$ assumes.
Hence, all good agents in $\textit{GC}$ can simulate $\texttt{PCONS}$.

Next, we prove this lemma.
By $g\geq 8f+8$, $(7/18)g\geq (7/18)(8f+8)=(56/18)f+56/18>3f$ holds.
Since all good agents in $\textit{GC}$ can simulate $\texttt{PCONS}$ and $(7/18)g>3f$ holds, by Lemma \ref{lem:PBCAlgorithm}, the executions of both $\texttt{PCONS}(a_i.S_p)$ and $\texttt{PCONS}(a_i.P_p)$ satisfy the PBC property and $a_i$ finishes both $\texttt{PCONS}(a_i.S_p)$ and $\texttt{PCONS}(a_i.P_p)$ in $O(f)$ cycles after cycle $c_i^\gamma$.
Thus, $a_i$ executes $a_i.\textit{stage}\leftarrow\textit{MakeGroup}$ in the last round of the cycle in which $a_i$ finishes both $\texttt{PCONS}(a_i.S_p)$ and $\texttt{PCONS}(a_i.P_p)$.
Hence, this lemma holds.
\end{proof}

For a good agent $a_i$, at the beginning of the second cycle of the $\textit{AgreeID}$ stage, $a_i.P_p$ contains all IDs of good agents in the same group candidate by Lemma \ref{lem:AtLeast7/18gGoodHaveAtLeast7/18gIDsInPp}, but does not contain IDs of the other good agents by the behavior of $\texttt{MakeReliableGroup}$.
By Lemma \ref{lem:AtLeast7/18gGoodCollectlyFinishAgreeID}, since the execution of $\texttt{PCONS}(a_i.P_p)$ satisfies the PBC property, by Validity 1 and Validity 2 of the PBC property, $a_i.P_c$ contains all IDs of good agents in the same group candidate but not IDs of the other good agents.
From this discussion, Lemma \ref{lem:AtLeast7/18gGoodCollectlyFinishAgreeID} and Corollary \ref{cor:GoodKnowAllGoodIds}, we have the following corollary.

\begin{corollary}
\label{cor:AgentsInGroupCandidateHaveTheSameSpAndPc}
Let $\textit{GC}$ be a group candidate, $a_i$ be a good agent in $\textit{GC}$, and $c_i^\gamma$ be the first cycle in which all good agents in $\textit{GC}$ are in the $\textit{MakeGroup}$ stage.
If at least $(7/18)g$ good agents belong to $\textit{GC}$, all good agents in $\textit{GC}$ have the same $P_c$ and $S_c$ at the beginning of cycle $c_i^\gamma$.
For each good agent $a_i$ in $\textit{GC}$, $|a_i.S_c|\geq g$ and $|a_i.P_c|\geq (7/18)g$ hold.
Also, $a_i.S_c$ contains all IDs of good agents, and $a_i.P_c$ contains all IDs of good agents in $\textit{GC}$ but not IDs of the other good agents.
\end{corollary}

Next, we consider the $\textit{MakeGroup}$ stage.
In the following two lemmas, we prove that when there exists at least one group candidate consisting of at least $(7/18)g$ good agents, at least one reliable group is created.

\begin{lemma}
\label{lem:7/18gGoodGatherAtSameNode}
Let $\textit{GC}$ be a group candidate, $a_i$ be a good agent in $\textit{GC}$, and $c_i^\gamma$ be a cycle in which all good agents in $\textit{GC}$ are in the $\textit{MakeGroup}$ stage.
If at least $(7/18)g$ good agents belong to $\textit{GC}$, $a_i$ stores a group ID in $a_i.\textit{gid}$ within $f+1$ cycles after cycle $c_i^\gamma$.
\end{lemma}
\begin{proof}
If $a_i$ has stored a group ID in $a_i.\textit{gid}$ before cycle $c_i^\gamma$, the lemma clearly holds.
Therefore, we consider the case where $a_i$ has not stored a group ID in $\textit{gid}$ before cycle $c_i^\gamma$.

First, we prove that all good agents in $\textit{GC}$ decide at least one same ID of a good agent in $\textit{GC}$ as a target ID within $f+1$ cycles after cycle $c_i^\gamma$.
By Corollary \ref{cor:AgentsInGroupCandidateHaveTheSameSpAndPc}, since at least $(7/18)g$ good agents belong to $\textit{GC}$, for each good agent $a_i$ in $\textit{GC}$, $|a_i.P_c|\geq (7/18)g$ holds at the beginning of cycle $c_i^\gamma$.
Since $(7/18)g\geq (7/18)(8f+8)>f+1$ holds by $g\geq 8f+8$, and agent $a_i$ in $\textit{GC}$ increments $a_i.\textit{count}$ by one in the last round of every cycle, $a_i$ uses different $f+1$ IDs in $a_i.P_c$ as target IDs during $f+1$ cycles starting from cycle $c_i^\gamma$.
Hence, since $f$ Byzantine agents exist in the network, the IDs contain at least one ID of a good agent in $a_i.P_c$.
Since all good agents in $\textit{GC}$ start the $\textit{AgreeID}$ stage at the same time by Definition \ref{def:GroupCandidate}, and they start a cycle at the same time by Observation \ref{obs:CountsOfConsensusAndGathringAreSame}, all good agents in $\textit{GC}$ have the same $\textit{count}$.
By Corollary \ref{cor:AgentsInGroupCandidateHaveTheSameSpAndPc}, all good agents in $\textit{GC}$ have the same $P_c$, and $a_i.P_c$ contains all IDs of good agents in $\textit{GC}$ but not IDs of the other good agents.
Thus, all good agents in $\textit{GC}$ decide the same agent ID in $P_c$ of them as a target ID in every cycle from cycle $c_i^\gamma$.
Hence, all good agents in $\textit{GC}$ decide at least one same ID of a good agent in $\textit{GC}$ as a target ID within $f+1$ cycles after cycle $c_i^\gamma$.

Next, let $c_i^\varepsilon$ be the first cycle that a good agent in $\textit{GC}$ has a target ID, $a_{gt}$ be the good agent that has the target ID in cycle $c_i^\varepsilon$, and $a_{gm}$ be the good agent with the largest ID of good agents in $\textit{GC}$.
We prove that all good agents in $\textit{GC}$ gather at a single node by the last round of cycle $c_i^\varepsilon$.
Agent $a_{gt}$ stays at the current node throughout cycle $c_i^\varepsilon$.
By the behavior of $\texttt{MakeReliableGroup}$, an agent searches for an agent with a target ID in the last half of a cycle.
Since $a_{gm}$ satisfies Line 8 of Algorithm \ref{alg:CollectIDStage}, and all good agents in $\textit{GC}$ have the same length of a cycle by Observation \ref{obs:CountsOfConsensusAndGathringAreSame}, $|c_i^\gamma|\geq 2\cdot (\TIME{REL}(a_{gm}.id)+1)$ holds.
On the other hand, by Lemma \ref{lem:RendezvousAlgorithm}, $a_i$ visits all nodes during $\TIME{REL}(a_i.id)$ rounds.
Therefore, $a_i$ meets $a_{gt}$ during the last half of cycle $c_i^\varepsilon$.
Hence, all good agents in $\textit{GC}$ gather at a single node by the last round of cycle $c_i^\varepsilon$.

Finally, we prove that $a_i$ stores a group ID in $a_i.\textit{gid}$ when all good agents in $\textit{GC}$ gather at a single node.
All good agents in $\textit{GC}$ have started the $\textit{MakeGroup}$ stage.
By Observation \ref{obs:CountsOfConsensusAndGathringAreSame}, all good agents in $\textit{GC}$ have the same $\textit{length}$.
By Corollary \ref{cor:AgentsInGroupCandidateHaveTheSameSpAndPc}, all good agents in $\textit{GC}$ have the same $S_c$ and, for each good agent $a_i$ in $\textit{GC}$, $|a_i.S_c|\geq g$ holds at the beginning of cycle $c_i^\gamma$.
Since $g>(8/9)k\geq (8/9)|a_i.S_p|$ holds by Corollary \ref{cor:g_geq_(8/9)k_geq_(8/9)Sp}, $|a_i.S_c|\geq g\geq (8/9)|a_i.S_p|$ holds.
For every good agent $a_j$ in $\textit{GC}$ other than $a_i$, since $g>(8/9)|a_i.S_p|$ and $|a_j.S_c|=|a_i.S_c|\geq g$ hold, $|a_j.S_c|\geq g\geq (8/9)|a_j.S_p|$ holds.
Therefore, $a_i.D$ contains at least IDs of all good agents in $\textit{GC}$.
Since $(7/18)g=(6/18)g+(1/18)g\geq (6/18)g+(1/18)(8f+8)=(6/18)g+(8/18)f+8/18>(6/18)(g+f)=(3/9)k$ holds by $g\geq 8f+8$, $|a_i.D|>(3/9)k\geq (3/9)|a_i.S_c|$ holds.
Thus, $a_i$ satisfies Line 13 of Algorithm \ref{alg:MakeGroupStage} and thus stores a group ID in $a_i.\textit{gid}$ when all good agents in $\textit{GC}$ gather at a single node.

Hence, this lemma holds.
\end{proof}

\begin{lemma}
\label{lem:WhenGoodAssignGroupIdToGid1/8kGoodExist}
Let $a_i$ be a good agent.
When $a_i$ stores a group ID in $a_i.\textit{gid}$, at least $k/8$ good agents store the same group ID as $a_i$ in their $\textit{gid}$.
\end{lemma}
\begin{proof}
First, we prove that $a_i.D$ includes at least $k/8$ IDs of good agents.
By the behavior of $\texttt{MakeReliableGroup}$, $a_i$ satisfies Line 13 of Algorithm \ref{alg:MakeGroupStage}.
Since $k\geq |a_i.S_p|\geq g$ holds by Corollary \ref{cor:GoodKnowAllGoodIds}, $|a_i.S_c|\geq (8/9)k\geq (8/9)|a_i.S_p|\geq (8/9)g$ holds.
Therefore, $|a_i.D|\geq (3/9)|a_i.S_c|\geq (3/9)(8/9)g=(8/27)g$ holds.
Since $f$ Byzantine agents exist in the network, at least $(8/27)g-f$ of them are good agents.
By $g\geq 8f+8$, $(8/27)g-f=(27/216)g+(37/216)g-f\geq (27/216)g+(37/216)(8f+8)-f=(27/216)g+(80/216)f+296/216>(27/216)(g+f)=(1/8)(g+f)=k/8$ holds.
Hence, $a_i.D$ includes at least $k/8$ IDs of good agents.

Next, we prove that at least $k/8$ good agents in $a_i.D$ store the same group ID as $a_i$ in their $\textit{gid}$ when $a_i$ stores a group ID in $a_i.\textit{gid}$.
Let $a_j$ be a good agent in $a_i.D$ other than $a_i$.
Since $a_i$ executes $a_i.D\leftarrow a_i.D\cup \{a_j.id\}$ by satisfying Line 12 of Algorithm \ref{alg:MakeGroupStage}, $|a_j.S_c|\geq (8/9)|a_j.S_p|$, $a_i.\textit{length}=a_j.\textit{length}$, $|a_i.S_c|=|a_j.S_c|$, and $a_j.\textit{stage}=\textit{MakeGroup}$ hold.
Since $a_i$ and $a_j$ observe the same states of agents at the current node, $a_i.D=a_j.D$ holds.
Since $|a_i.D|\geq (3/9)|a_i.S_c|$, $a_i.D=a_j.D$, and $|a_i.S_c|=|a_j.S_c|$ hold, $|a_j.D|\geq (3/9)|a_j.S_c|$ holds.
Therefore, $a_j$ satisfies Line 13 of Algorithm \ref{alg:MakeGroupStage} and thus stores the same group ID in $a_j.\textit{gid}$.

Hence, this lemma holds.
\end{proof}

Finally, we prove the complexity of $\texttt{MakeReliableGroup}$.

\begin{theorem}
\label{the:MakeReliableGroupConstruction}
Let $n$ be the number of nodes, $k$ be the number of agents, $g$ be the number of good agents, $f$ be the number of weakly Byzantine agents, $a_{max}$ be a good agent with the largest ID among good agents.
If the upper bound $N$ of $n$ is given to agents and $k\geq 9f+8$ holds, Algorithm \ref{alg:MakeReliableGroupConstruction} makes good agents create at least one reliable group in $O(f\cdot\TIME{REL}(a_{max}.id))$ rounds.
\end{theorem}
\begin{proof}
By Lemma \ref{lem:AllGoodFinishMCInOt_REL}, all good agents finish the $\textit{MakeCandidate}$ stage in $O(\TIME{REL}(a_{max}.id))$ rounds after starting $\texttt{MakeReliableGroup}$.
By Lemma \ref{lem:AtLeast7/9gGoodTransitToAgreeIDInEitherOfTheTwoRounds} and Corollary \ref{cor:7/18gGoodStartAgreeIDTogether}, in $O(\TIME{REL}(a_{max}.id))$ rounds after starting $\texttt{MakeReliableGroup}$, there is at least one round that at least $(7/18)g$ good agents start the $\textit{AgreeID}$ stage at the same time.
Let $\textit{GC}$ be a group candidate of at least $(7/18)g$ good agents, $c_{gc}^\zeta$ be a cycle that all good agents in $\textit{GC}$ start the $\textit{AgreeID}$ stage, and $c_{gc}^\eta$ be the first cycle in which all good agents in $\textit{GC}$ are in the $\textit{MakeGroup}$ stage.
By Lemma \ref{lem:AtLeast7/18gGoodCollectlyFinishAgreeID}, all good agents in $\textit{GC}$ have started the $\textit{MakeGroup}$ stage in $O(f)$ cycles after cycle $c_{gc}^\zeta$.
By Lemma \ref{lem:7/18gGoodGatherAtSameNode}, a good agent in $\textit{GC}$ stores a group ID in its $\textit{gid}$ within $f+1$ cycles after cycle $c_i^\eta$. 
Also, by Lemma \ref{lem:WhenGoodAssignGroupIdToGid1/8kGoodExist}, when a good agent in $\textit{GC}$ stores a group ID in its $\textit{gid}$, at least $k/8$ good agents store the same group ID as the agent in their $\textit{gid}$. 
That is, a reliable group is created within $f+1$ cycles after cycle $c_i^\eta$.
Therefore, since the maximum length of the cycles is at most $32\cdot (\TIME{REL}(a_{max}.id)+1)$ by Lemma \ref{lem:MaxLength}, a reliable group is created in $32\cdot (\TIME{REL}(a_{max}.id)+1)\cdot O(f)+32\cdot (\TIME{REL}(a_{max}.id)+1)(f+1)=O(f\cdot\TIME{REL}(a_{max}.id))$ rounds after starting cycle $c_{gc}^\zeta$. 
Hence, since the first round of cycle $c_{gc}^\zeta$ is within $O(\TIME{REL}(a_{max}.id))$ rounds after starting $\texttt{MakeReliableGroup}$, a reliable group is created in $O(\TIME{REL}(a_{max}.id))+O(f\cdot\TIME{REL}(a_{max}.id))=O(f\cdot\TIME{REL}(a_{max}.id))$ after starting $\texttt{MakeReliableGroup}$.
Hence, this theorem holds.
\end{proof}

\subsection{Gathering Algorithm}
\label{sec:GatheringAlgorithm}
In this section, we propose an algorithm that solves the gathering problem by assuming that $k=g+f\geq 9f+8$.
The proposed algorithm uses $\texttt{MakeReliableGroup}$ described in Section \ref{sec:MakeReliableGroupConstruction} to create at least one reliable group.
Recall that agents know $N$, but do not know $n,k,f$ or $F$.

\subsubsection{Description of the Algorithm}
\begin{algorithm}[t]
  \caption{$\texttt{ByzantineGathering}(N)$ for agent $a_i$}
  \label{alg:ByzantineGathering}
  \begin{algorithmic}[1]
    \If{$a_i.\textit{stage}=\textit{CollectID}$}
      \State Execute $\texttt{MakeReliableGroup}$
    \Else \Comment{$a_i.\textit{stage}\in \{\textit{MakeCandidate}$, $\textit{AgreeID}$, $\textit{MakeGroup}\}$}
      \State $a_i.S_{gid}\leftarrow \{x\mid \exists A_{rg}\subset A_i[|A_{rg}|\geq (1/8)|a_i.S_p|\wedge \forall a_j\in A_{rg}:a_j.\textit{gid}=x]\}$
      \If{$a_i.S_{gid}\neq \emptyset$}
        \State $a_i.\textit{minGID}\leftarrow\min(a_i.S_{gid})$
      \EndIf
      \If{$a_i.S_{gid}\neq \emptyset\wedge a_i.\textit{gid}>a_i.\textit{minGID}$}
        \State $a_i.S_{rg}\leftarrow\{id\mid \exists a_j\in A_i[a_j.\textit{gid}=a_i.\textit{minGID}\wedge a_j.id=id]\}$
        \State Execute $\texttt{FOLLOW}(a_i.S_{rg})$
      \ElsIf{$a_i.\textit{gid}\neq \infty$}
        \State $a_i.\textit{elapsed}\leftarrow a_i.\textit{elapsed}+1$
        \If{$a_i.\textit{length}=a_i.\textit{elapsed}$}
          \State Execute \texttt{TERMINATE}()
        \EndIf
        \State Execute $\texttt{REL}(a_i.\textit{gid})(a_i.\textit{elapsed})$
      \Else
        \State Execute $\texttt{MakeReliableGroup}$
      \EndIf
    \EndIf
  \end{algorithmic} 
\end{algorithm}

Algorithm \ref{alg:ByzantineGathering} shows the behavior of each round of Algorithm $\texttt{ByzantineGathering}$.
The proposed algorithm aims to make all good agents transition into the terminal state at the same node using a reliable group.
In $\texttt{ByzantineGathering}$, we introduce procedures $\texttt{TERMINATE}()$ and $\texttt{FOLLOW}(S_{rg})$.
Procedure $\texttt{TERMINATE}()$ means that an agent transitions into the terminal state.
Procedure $\texttt{FOLLOW}(S_{rg})$ means that an agent $a_i$ executes the following two actions:
(1) When a majority of agents in $S_{rg}$ move to some node, $a_i$ also moves to the node.
(2) When a majority of agents in $S_{rg}$ execute $\texttt{TERMINATE}()$ or have entered a terminal state, $a_i$ executes $\texttt{TERMINATE}()$.
Agent $a_i$ can refer to the variables used in $\texttt{MakeReliableGroup}$ at any time.

First, we provide the overall execution of $\texttt{ByzantineGathering}$.
Every good agent executes $\texttt{MakeReliableGroup}$ unless it does not stay with a reliable group at a node.
After creating a reliable group, good agents in the group collect the other good agents.
To do this, when agents create a reliable group in round $r$, a good agent $a_\ell$ in the reliable group executes $\texttt{REL}(a_\ell.\textit{gid})$ for $a_\ell.\textit{length}$ rounds after round $r+1$.
Then, when a good agent meets the reliable group with a smaller group ID, it accompanies the group.
By Lemma \ref{lem:WhenGoodAssignGroupIdToGid1/8kGoodExist}, when a good agent stores a group ID in its $\textit{gid}$, at least $k/8$ good agents store the same group ID in their $\textit{gid}$.
Thus, a reliable group contains at least $k/8$ good agents.
Since $k\geq |a_i.S_p|\geq g$ holds by Corollary \ref{cor:GoodKnowAllGoodIds}, $k/8\geq (1/8)|a_i.S_p|\geq g/8$ holds.
Therefore, if a reliable group contains at least one good agent and exists at the node with $a_i$, $a_i$ can recognize the group.
On the other hand, by $g\geq 8f+8$, $(1/8)|a_i.S_p|\geq g/8\geq f+1$ holds and thus only $f$ Byzantine agents are not enough to create a reliable group.
Thus, if a group of only $f$ Byzantine agents exists at the node with $a_i$, $a_i$ does not recognize the group as a reliable group.
Summarizing the above, we have the following observation.

\begin{observation}
\label{obs:GoodRecognizeReliableGroup}
If a good agent meets a reliable group consisting of at least $k/8$ good agents with the same group ID, the agent recognizes the group as a reliable group, otherwise, the agent does not recognize the group as a reliable group.
\end{observation}

Furthermore, a reliable group contains at least $k/8$ good agents, by $k\geq 9f+8$, $k/8\geq (9/8)f+1>f+1$ holds.
That is, a reliable group contains more than $f+1$ good agents.
Thus, good agents are the majority of agents in a reliable group even if the group contains $f$ Byzantine agents.
Therefore, when a good agent meets a reliable group, it can trust the group.
By the behavior of $\texttt{MakeReliableGroup}$, round $r$ is the last round of some cycle of the $\textit{MakeGroup}$ stage, and thus round $r+1$ is the first round of the next cycle of that.
All good agents in the same reliable group have the same length of their cycles, and thus the execution of $\texttt{REL}(a_\ell.\textit{gid})$ can be regarded as the execution of an additional cycle of the $\textit{MakeGroup}$ stage by $a_\ell$.
The following observation summarizes this discussion.

\begin{observation}
\label{obs:RGCycleIsIdenticalToCycleMRG}
When a good agent $a_i$ belongs to a reliable group, the start round and length of the execution of $\texttt{REL}(a_i.\textit{gid})$ are the same as those of the last cycle in the $\textit{MakeGroup}$ stage of $a_i$.
\end{observation}

Therefore, by Lemma \ref{lem:GoodMeetAllGood}, when $a_\ell$ executes $\texttt{REL}(a_\ell.\textit{gid})$ for $a_\ell.\textit{length}$ rounds after round $r+1$, $a_\ell$ meets all good agents not in the same reliable group.
Consequently, all good agents accompany the reliable group with the smallest ID and achieve the gathering.

Hereinafter, we explain the detailed behaviors of $\texttt{ByzantineGathering}$.
At the beginning of every round, an agent $a_i$ determines its action depending on the states of $a_i$ and other agents at the current node.

If $a_i.\textit{stage}=\textit{CollectID}$ holds, it executes $\texttt{MakeReliableGroup}$ until it collects IDs of all good agents (Lines 1--2 of Algorithm \ref{alg:ByzantineGathering}).

If $a_i.\textit{stage}\in \{\textit{MakeCandidate}$, $\textit{AgreeID}$, $\textit{MakeGroup}\}$ holds, $a_i$ determines whether the current node contains a reliable group or not.
First, if $a_i$ finds groups with at least $(1/8)|a_i.S_p|$ agents, it stores the group IDs to variable $a_i.S_{gid}$ (Line 4). 
After that, if $a_i.S_{gid}$ contains at least one group ID, $a_i$ calculates the smallest group ID among $a_i.S_{gid}$ and stores the smallest ID in variable $\textit{minGID}$ (Line 5--7).
Initially, $a_i.S_{gid}=\emptyset$ and $a_i.\textit{minGID}=\infty$ hold.

Consider the case where $a_i.S_{gid}$ contains at least one group ID and $a_i.\textit{minGID}$ is smaller than $a_i.\textit{gid}$ (Line 8).
This implies one of the following two conditions.
\begin{itemize}
  \item[(1)] Agent $a_i$ is not a member of a reliable group and meets a reliable group at the current node.
  \item[(2)] Agent $a_i$ is a member of a reliable group and meets a reliable group with a group ID smaller than $a_i.\textit{gid}$ at the current node.
\end{itemize}
In this case, $a_i$ finds agents with $\textit{gid}=a_i.\textit{minGID}$ and stores their IDs in variable $a_i.S_{rg}$.
Then, $a_i$ follows the action of the majority of agents in $a_i.S_{rg}$ (Lines 8--10).

If $a_i$ is a member of a reliable group with the smallest group ID at the current node, it executes $\texttt{REL}(a_i.\textit{gid})$ to meet other good agents and then transitions into a terminal state (Lines 11--16).
Note that, if $a_i$ meets a reliable group with a group ID smaller than $a_i.\textit{gid}$ during the execution, it follows the group as in the previous paragraph.
When $a_i$ executes $\texttt{REL}(a_i.\textit{gid})$ for $a_i.\textit{length}$ rounds, by Observation \ref{obs:RGCycleIsIdenticalToCycleMRG}, the execution and a cycle of the $\textit{MakeGroup}$ stage in $\texttt{MakeReliableGroup}$ start and finish at the same time.
Thus, by Lemma \ref{lem:GoodMeetAllGood}, $a_i$ meets all good agents.

If $a_i$ does not satisfy the above conditions (i.e., the current node does not contain a reliable group), it executes $\texttt{MakeReliableGroup}$ for one round (Lines 17--19).

\subsubsection{Correctness and Complexity Analysis}
In this subsection, we prove the correctness and complexity of the proposed algorithm.

By the behavior of $\texttt{ByzantineGathering}$, an agent executes $\texttt{MakeReliableGroup}$ until the current node contains a reliable group.
By Theorem \ref{the:MakeReliableGroupConstruction}, agents create at least one reliable group after starting $\texttt{ByzantineGathering}$.
Therefore, we prove that, after a reliable group is created, all good agents gather at a single node using a reliable group and transition into a terminal state.

Let $r_{ini}$ be the round in which the first reliable group is created, $\textit{SRG}_{ini}$ be a set of the reliable groups created in round $r_{ini}$, and $\textit{RG}_{min}$ be the reliable group with the smallest group ID among $\textit{SRG}_{ini}$.
Let $A_{ec}$ be a set of good agents that execute the $\textit{CollectID}$ stage in round $r_{ini}+1$.
Let $A_{fc}$ be a set of good agents that have finished the $\textit{CollectID}$ stage and do not belong to a reliable group in round $r_{ini}+1$.
For a good agent $a_i\in A_{ec}\cup A_{fc}$, let $c_i^*$ be a cycle of $a_i$ that includes round $r_{ini}+1$ and round $r^*_i$ be the last round of cycle $c^*_i$.
For a good agent $a_j$ in some reliable group of $\textit{SRG}_{ini}$, let round $r^*_j$ be the last round of the execution of $\texttt{REL}(a_j.\textit{gid})$.
Let $r^*=\max(\{r^*_i\mid a_i$ is a good agent in $\textit{MA}\})$.
By the behavior of $\texttt{MakeReliableGroup}$, since an agent extends the length of its cycle in the $\textit{CollectID}$ and $\textit{MakeCandidate}$ stages but not in the $\textit{AgreeID}$ and $\textit{MakeGroup}$ stages, good agents in the $\textit{CollectID}$ and $\textit{MakeCandidate}$ stages have the longest cycles.
By Observation \ref{obs:RGCycleIsIdenticalToCycleMRG}, when a good agent $a_j$ belongs to a reliable group in $\textit{SRG}_{ini}$, the length of the execution of $\texttt{REL}(a_j.\textit{gid})$ is the same as that of the last cycle in the $\textit{MakeGroup}$ stage of $a_j$.
Thus, by Observation \ref{obs:LargerXGoodStartTimeAndSmallerXGoodStartTimeSame}, when good agents in the $\textit{CollectID}$ and $\textit{MakeCandidate}$ stages start a cycle, $a_j$ and good agents in the $\textit{AgreeID}$ and $\textit{MakeGroup}$ stages also start executing $\texttt{REL}(a_j.\textit{gid})$ and a cycle.
Therefore, the period of a cycle of good agents in the $\textit{CollectID}$ and $\textit{MakeCandidate}$ stages includes the period of a cycle of good agents in the $\textit{AgreeID}$ and $\textit{MakeGroup}$ stages and the execution of $\texttt{REL}(a_j.\textit{gid})$.
By Observation \ref{obs:XsOfCollectIDandPreparationAreSame}, good agents in the $\textit{CollectID}$ and $\textit{MakeCandidate}$ stages have the same length of their cycles.
Hence, if $A_{ec}$ includes at least one good agent, $r^*$ is the last round of the cycle of an agent in $A_{ec}$.
From this discussion, we have the following observation.

\begin{observation}
\label{obs:r^startIsTheLastRoundOfCycleOfAec}
If $A_{ec}$ includes at least one good agent, $r^*$ is the last round of cycle $c^*_i$ for some agent $a_i\in A_{ec}$.
\end{observation}

In the following lemma, we prove that a good agent $a_i$ in $A_{fc}$ and $\textit{RG}_{min}$ meet if $a_i$ and good agents in $\textit{RG}_{min}$ never interrupt $\texttt{MakeReliableGroup}$ and $\texttt{REL}$ with the group ID in any round before they meet.

\begin{lemma}
\label{lem:GoodsNotInReliableGroupMeetReliableGroup}
Assume that a good agent $a_i$ in $A_{fc}$ executes $\texttt{MakeReliableGroup}$ until the last round of cycle $c^*_i$, and every good agent $a_j$ in $\textit{RG}_{min}$ executes $\texttt{REL}(a_j.\textit{gid})$ for $a_j.\textit{length}$ rounds without interruption after round $r_{ini}+1$.
In this case, $a_i$ and $\textit{RG}_{min}$ meet before the last round of cycle $c^*_i$.
\end{lemma}
\begin{proof}
First, we consider the case that the first round of cycle $c^*_i$ is round $r_{ini}+1$.
In this case, by the behavior of $\texttt{MakeReliableGroup}$, $a_i$ executes $\texttt{REL}(a_i.id)$ for at least the first half of cycle $c^*_i$.
Since $a_i$ and $a_j$ satisfy Line 8 of Algorithm \ref{alg:CollectIDStage} before round $r_{ini}$, $|c^*_i|\geq 2\cdot (\TIME{REL}(a_i.id)+1)$ and $a_j.\textit{length}\geq 2\cdot (\TIME{REL}(a_j.id)+1)$ hold.
Since $a_j.id\geq a_j.\textit{gid}$ holds by the behavior of $\texttt{MakeReliableGroup}$, $a_j.\textit{length}\geq 2\cdot (\TIME{REL}(a_j.\textit{gid})+1)$ holds.
Thus, $a_i$ and $a_j$ execute $\texttt{REL}(a_i.id)$ and $\texttt{REL}(a_j.\textit{gid})$ at the same time for at least $\TIME{REL}(\min(a_i.id,a_j.\textit{gid}))$ rounds.
Also, by Definition \ref{def:ReliableGroup}, all good agents in $\textit{RG}_{min}$ have the same $\textit{gid}$ and $\textit{length}$.
Thus, they behave identically during the execution of $\texttt{REL}(a_j.\textit{gid})$.
Therefore, by Lemma \ref{lem:RendezvousAlgorithm}, $a_i$ meets $\textit{RG}_{min}$ in at least $\TIME{REL}(\min(a_i.id,a_j.\textit{gid}))$ rounds after round $r_{ini}+1$, that is, before the last round of cycle $c^*_i$.

Next, we consider the case that the first round of cycle $c^*_i$ is not round $r_{ini}+1$.
By Observations \ref{obs:LargerXGoodStartTimeAndSmallerXGoodStartTimeSame} and \ref{obs:RGCycleIsIdenticalToCycleMRG}, if $|c^*_i|\leq a_j.\textit{length}$ holds, when $a_j$ starts executing $\texttt{REL}(a_j.\textit{gid})$, $a_i$ also starts cycle $c^*_i$.
That is, this case does not apply if $|c^*_i|\leq a_j.\textit{length}$ holds.
Thus, this case implies that $|c^*_i|>a_j.\textit{length}$ holds.
By Observations \ref{obs:LargerXGoodStartTimeAndSmallerXGoodStartTimeSame} and \ref{obs:RGCycleIsIdenticalToCycleMRG}, the execution of $\texttt{REL}(a_j.\textit{gid})$ is completely included in the period of cycle $c^*_i$.
Therefore, by Lemma \ref{lem:GoodMeetAllGood}, $a_j$ meets $a_i$ by the last round of execution of $\texttt{REL}(a_j.\textit{gid})$.
Since all good agents in $\textit{RG}_{min}$ behave identically during the execution of $\texttt{REL}(a_j.\textit{gid})$, $a_i$ and $\textit{RG}_{min}$ meet before the last round of the execution of $\texttt{REL}(a_j.\textit{gid})$.
\end{proof}

For an agent $a_i$ and a reliable group $\textit{RG}$, we say ``$a_i$ follows $\textit{RG}$ in round $r$'' if $a_i$ and $\textit{RG}$ satisfy the following condition: (1) If all good agents in $\textit{RG}$ terminate in round $r$, $a_i$ also terminates in round $r$, and (2) If all good agents in $\textit{RG}$ move to node $v$, $a_i$ also moves to node $v$.
In the following lemma, we prove that, if a good agent meets a reliable group, the good agent follows the reliable group with the smallest group ID at the node.

\begin{lemma}
\label{lem:WhenGoodMeetRGMinTheyFollow}
Assume that at least one reliable group exists at a node $v$ in round $r'$.
Let $\textit{RG}'$ be the reliable group with the smallest group ID at $v$ in round $r'$.
Then, all good agents not in $\textit{RG}'$ at $v$ follow $\textit{RG}'$ in round $r'$.
\end{lemma}
\begin{proof}
Let $a_i$ be a good agent not in $\textit{RG}'$ at $v$ in round $r'$.
Let $C'$ be a set of all group IDs of reliable groups at $v$ in round $r'$.
By Observation \ref{obs:GoodRecognizeReliableGroup}, $a_i$ recognizes all reliable groups at $v$.
Thus, $a_i$ executes $a_i.S_{gid}\leftarrow C'$, and then $a_i$ executes $a_i.\textit{minGID}\leftarrow \min(a_i.S_{gid})$.
This implies that $a_i$ executes $a_i.\textit{minGID}\leftarrow \min(C')$.
By the assumption of this lemma, since $a_i.S_{gid}\neq \emptyset$ and $a_i.\textit{gid}>a_i.\textit{minGID}$ hold in round $r'_i$, $a_i$ calculates $a_i.S_{rg}$ and executes $\texttt{FOLLOW}(a_i.S_{rg})$.
Note that $a_i.S_{rg}$ contains all good agents in $\textit{RG}'$.
Recall that all good agents in $\textit{RG}'$ make the same behavior and the number of good agents in $\textit{RG}'$ is at least $f+1$ by Definition \ref{def:ReliableGroup}, Lemma \ref{lem:WhenGoodAssignGroupIdToGid1/8kGoodExist}, and $k\geq 9f+8$.
Hence, $a_i$ follows $\textit{RG}'$.
\end{proof}

In the following lemma, we prove that agents do not create a reliable group between round $r_{ini}+1$ and round $r^*$ inclusive.

\begin{lemma}
\label{lem:RoundInWhichReliableGroupsAreMadeAreOnlyOne}
Assume that every good agent $a_j$ in $\textit{RG}_{min}$ executes $\texttt{REL}(a_j.\textit{gid})$ for $a_j.\textit{length}$ rounds without interruption after round $r_{ini}+1$.
No reliable group is created between round $r_{ini}+1$ and round $r^*$ inclusive.
\end{lemma}
\begin{proof}
We prove this lemma by contradiction.
Assume that a reliable group is created in round $r'\ (r^*\geq r'\geq r_{ini}+1)$ and no reliable group is created between round $r_{ini}+1$ and round $r'-1$ inclusive.
By Observation \ref{obs:r^startIsTheLastRoundOfCycleOfAec}, if $A_{ec}$ includes at least one good agent, an agent in $A_{ec}$ starts the $\textit{MakeCandidate}$ stage after round $r^*$ and hence it cannot participate in the creation of a reliable group.
Thus, it is sufficient to consider only agents in $A_{fc}$.
By the behavior of $\texttt{ByzantineGathering}$, each agent $a_i\in A_{fc}$ behaves according to $\texttt{MakeReliableGroup}$ until the current node contains a reliable group.
On the other hand, by Lemma \ref{lem:WhenGoodMeetRGMinTheyFollow}, if $a_i$ meets a reliable group, it follows the reliable group and hence cannot participate in the creation of a reliable group.
Therefore, to derive a contradiction, it is enough to prove that $a_i$ meets some reliable group before round $r^*$.
However, if $a_i$ continues $\texttt{MakeReliableGroup}$ without meeting a reliable group, by Lemma \ref{lem:GoodsNotInReliableGroupMeetReliableGroup}, $a_i$ and $\textit{RG}_{min}$ meet before the last round of cycle $c^*_i$.
Since round $r'$ is the last round of cycle $c^*_i$ or later by the behavior of $\texttt{MakeReliableGroup}$, and $r^*\geq r'$ holds, $a_i$ meets $\textit{RG}_{min}$ before round $r^*$.
This is a contradiction.
Hence, this lemma holds.
\end{proof}

In the following lemma, we prove that good agents in $\textit{RG}_{min}$ never interrupt $\texttt{REL}$.

\begin{lemma}
\label{lem:RGMinDoesNotStop}
Every good agent $a_i$ in $\textit{RG}_{min}$ executes $\texttt{REL}(a_i.\textit{gid})$ for $a_i.\textit{length}$ rounds without interruption after round $r_{ini}+1$ and then transitions into a terminal state.
\end{lemma}
\begin{proof}
By the behavior of $\texttt{ByzantineGathering}$, $a_i$ interrupts $\texttt{REL}(a_i.\textit{gid})$ if it meets a reliable group with a smaller group ID.
However, by Lemma \ref{lem:RoundInWhichReliableGroupsAreMadeAreOnlyOne}, no reliable group is created between round $r_{ini}+1$ and round $r^*$ inclusive, and hence the group ID of $\textit{RG}_{min}$ is the smallest until round $r^*$.
This implies that, since $r^*\geq r_{ini}+a_i.\textit{length}$ holds, $a_i$ executes $\texttt{REL}(a_i.\textit{gid})$ for $a_i.\textit{length}$ rounds without interruption.
\end{proof}

In the following lemma, we prove that two reliable groups in $\textit{SRG}_{ini}$ meet.

\begin{lemma}
\label{lem:AllRGMeet}
Let $\textit{RG}$ be a reliable group in $\textit{SRG}_{ini}\setminus\{\textit{RG}_{min}\}$.
Then, all good agents in $\textit{RG}$ meet a reliable group with a group ID smaller than $\textit{RG}$ before round $r^*$.
\end{lemma}
\begin{proof}
Let $a_i$ be an arbitrary good agent in $\textit{RG}_{min}$.
By Lemma \ref{lem:RGMinDoesNotStop}, $a_i$ executes $\texttt{REL}(a_i.\textit{gid})$ for $a_i.\textit{length}$ rounds without interruption after round $r_{ini}+1$.

For contradiction, assume that some agent $a_j$ in $\textit{RG}$ never meets a reliable group with a group ID smaller than $\textit{RG}$.
In this case, $a_j$ executes $\texttt{REL}(a_j.\textit{gid})$ for $a_j.\textit{length}$ rounds without interruption after round $r_{ini}+1$.
Also, $a_i.id\geq a_i.\textit{gid}$ and $a_j.id\geq a_j.\textit{gid}$ hold.
Thus, since $a_i$ and $a_j$ satisfy Line 8 of Algorithm \ref{alg:CollectIDStage}, $a_i.\textit{length}\geq 2\cdot(\TIME{REL}(a_i.id)+1)\geq 2\cdot(\TIME{REL}(a_i.\textit{gid})+1)$ and $a_j.\textit{length}\geq 2\cdot(\TIME{REL}(a_j.id)+1)\geq 2\cdot(\TIME{REL}(a_j.\textit{gid})+1)$ hold.
Therefore, since $a_i.\textit{gid}<a_j.\textit{gid}$ holds, by the behavior of $\texttt{ByzantineGathering}$, $a_i$ and $a_j$ execute $\texttt{REL}(a_i.\textit{gid})$ and $\texttt{REL}(a_j.\textit{gid})$ at the same time for at least $\TIME{REL}(a_i.\textit{gid})$ rounds after round $r_{ini}+1$.
By Lemma \ref{lem:RendezvousAlgorithm}, $a_i$ and $a_j$ meet in $\TIME{REL}(a_i.\textit{gid})$ rounds after round $r_{ini}+1$.
Since $r^*>r_{ini}+\TIME{REL}(a_i.\textit{gid})$ holds and all good agents in $\textit{RG}_{min}$ stay at the same node, $a_j$ meets $\textit{RG}_{min}$.
This is a contradiction.
\end{proof}

\begin{lemma}
\label{lem:WhenGoodFinishCITheyGather}
All good agents in $A_{fc}$ and all good agents in reliable groups in $\textit{SRG}_{ini}$ gather at the node with $\textit{RG}_{min}$ and transition into a terminal state by round $r^*$.
\end{lemma}
\begin{proof}
By Lemma \ref{lem:RGMinDoesNotStop}, every good agent $a_i$ in $\textit{RG}_{min}$ executes $\texttt{REL}(a_i.\textit{gid})$ without interruption and transitions into a terminal state before round $r^*$.
In addition, if a good agent meets $\textit{RG}_{min}$, it follows $\textit{RG}_{min}$ after that by Lemma \ref{lem:WhenGoodMeetRGMinTheyFollow}.
Hence, it is sufficient to prove that all good agents in $A_{fc}$ and all good agents in reliable groups other than $\textit{RG}_{min}$ meet $\textit{RG}_{min}$ before round $r^*$.

Consider a reliable group $\textit{RG}$ in $\textit{SRG}_{ini}\setminus\{\textit{RG}_{min}\}$.
By Lemma \ref{lem:AllRGMeet}, agents in $\textit{RG}$ meet a reliable group $\textit{RG}'$ with a smaller group ID before round $r^*$.
If $\textit{RG}$ is $\textit{RG}_{min}$, agents in $\textit{RG}$ meet $\textit{RG}_{min}$.
Otherwise, good agents in $\textit{RG}$ follow $\textit{RG}'$ by Lemma \ref{lem:WhenGoodMeetRGMinTheyFollow}.
Similarly good agents in $\textit{RG}'$ meet another reliable group $\textit{RG}''$ with a smaller group ID before $r^*$.
Hence, good agents in $\textit{RG}$ also meet $\textit{RG}''$.
By repeating this discussion, eventually good agents in $\textit{RG}$ meet $\textit{RG}_{min}$ before round $r^*$.

Lastly consider a good agent $a_i$ in $A_{fc}$.
As long as $a_i$ does not meet a reliable group, it continues $\texttt{MakeReliableGroup}$.
In this case, by Lemma \ref{lem:GoodsNotInReliableGroupMeetReliableGroup}, $a_i$ meets $\textit{RG}_{min}$ before round $r^*$.
Otherwise, if $a_i$ meets a reliable group, it follows the reliable group by Lemma \ref{lem:WhenGoodMeetRGMinTheyFollow}.
In this case, similarly to the previous paragraph, $a_i$ meets $\textit{RG}_{min}$ before round $r^*$.
\end{proof}

Finally, we prove the correctness and complexity of $\texttt{ByzantineGathering}$.

\begin{theorem}
\label{the:ByzantineGathering}
Let $n$ be the number of nodes, $k$ be the number of agents, $f$ be the number of weakly Byzantine agents, and $a_{max}$ be a good agent with the largest ID among good agents.
If the upper bound $N$ of $n$ is given to agents and $k\geq 9f+8$ holds, Algorithm \ref{alg:ByzantineGathering} solves the gathering problem in $O(f\cdot\TIME{REL}(a_{max}.id))$ rounds.
\end{theorem}
\begin{proof}
By the behavior of $\texttt{ByzantineGathering}$, an agent executes $\texttt{MakeReliableGroup}$ until the current node contains a reliable group.
By Theorem \ref{the:MakeReliableGroupConstruction}, a reliable group is created in $O(f\cdot\TIME{REL}(a_{max}))$ rounds after starting $\texttt{MakeReliableGroup}$.
Thus, round $r_{ini}$ is in $O(f\cdot\TIME{REL}(a_{max}))$ rounds after starting $\texttt{MakeReliableGroup}$. 

First, we consider agents in $A_{fc}$ or some reliable group of $\textit{SRG}_{ini}\setminus\{\textit{RG}_{min}\}$.
By Lemma \ref{lem:WhenGoodFinishCITheyGather}, all good agents in $A_{fc}$ or some reliable group of $\textit{SRG}_{ini}\setminus\{\textit{RG}_{min}\}$ gather at the node with $\textit{RG}_{min}$ and transition into a terminal state by round $r^*$.

Next, we consider agents in $A_{ec}$.
Let $a_\ell$ be a good agent in $A_{ec}$.
By Observation \ref{obs:r^startIsTheLastRoundOfCycleOfAec}, round $r^*$ is the last round of cycle $c^*_\ell$.
Therefore, from discussion in the previous paragraph, when $a_\ell$ starts the $\textit{MakeCandidate}$ stage, all good agents in $A_{fc}$ or some reliable group of $\textit{SRG}_{ini}$ have gathered with $\textit{RG}_{min}$ and entered a terminal state.
Also, since $a_\ell$ satisfies Line 8 of Algorithm \ref{alg:CollectIDStage}, by the behavior of $\texttt{MakeReliableGroup}$, $a_\ell$ executes $\texttt{REL}(a_\ell.id)$ for at least $\TIME{REL}(a_\ell.id)$ rounds in the first cycle of the $\textit{MakeCandidate}$ stage.
Therefore, by Lemma \ref{lem:RendezvousAlgorithm}, $a_\ell$ visits all nodes in the first cycle of the $\textit{MakeCandidate}$ stage and thus meets $\textit{RG}_{min}$ during the cycle.
By Lemma \ref{lem:WhenGoodMeetRGMinTheyFollow}, $a_\ell$ follows $\textit{RG}_{min}$ at that time and hence transitions into a terminal state by the last round of the first cycle of the $\textit{MakeCandidate}$.

Finally, we analyze the time complexity of $\texttt{ByzantineGathering}$.
From the above discussion, the time required to achieve the gathering depends on whether $A_{ec}$ contains at least one good agent or not.
Thus, since $a_{max}$ finishes the $\textit{CollectID}$ stage the latest, we consider the two cases, one where $A_{ec}$ does not contain $a_{max}$ and the other where $A_{ec}$ contains $a_{max}$.
In the former case, since $a_i.\textit{length}<32\cdot (\TIME{REL}(a_{max}.id)+1)$ holds for any good agent $a_i$ by Lemma \ref{lem:MaxLength}, $r^*\leq r_{ini}+32\cdot (\TIME{REL}(a_{max}.id)+1)+1$ holds.
Thus, good agents achieve the gathering in $r_{ini}+a_i.\textit{length}+1<O(f\cdot\TIME{REL}(a_{max}))+32\cdot (\TIME{REL}(a_{max}.id)+1)+1=O(f\cdot\TIME{REL}(a_{max}))$ rounds.
In the latter case, $r^*$ is the last round of cycle $c^*_{max}$ by Observation \ref{obs:r^startIsTheLastRoundOfCycleOfAec}.
Also, $a_{max}$ finishes the $\textit{CollectID}$ stage in $O(\TIME{REL}(a_{max}.id))$ rounds after starting $\texttt{MakeReliableGroup}$ by Lemma \ref{lem:AllGoodFinishMCInOt_REL}.
Thus, since $a_{max}.\textit{length}<32\cdot (\TIME{REL}(a_{max}.id)+1)$ holds, good agents achieve the gathering in $O(\TIME{REL}(a_{max}.id))+32\cdot (\TIME{REL}(a_{max}.id)+1)=O(\TIME{REL}(a_{max}.id))$ rounds.

Hence, all good agents gather at the same node and transition into a terminal state in $O(f\cdot\TIME{REL}(a_{max}))$ rounds after starting $\texttt{ByzantineGathering}$.
\end{proof}

\section{Conclusion}
In this paper, we have developed an algorithm that achieves the gathering with non-simultaneous termination in weakly Byzantine environments.
The algorithm reduces time complexity compared to the existing algorithm if $n$ is given to agents, although the guarantees on simultaneous termination and startup delay are not the same. %
More specifically, the proposed algorithm achieves the gathering in $O(f\cdot |\Lambda_{good}|\cdot X(N))$ rounds if the upper bound $N$ of the number of nodes is given to agents and at least $9f+8$ agents exist in the network.
In the algorithm, several good agents first create a reliable group so that good agents can trust the behavior of the group to suppress the influence of Byzantine agents.
After that, the reliable group collects the other good agents, and all good agents gather at a single node.
To create a reliable group, several good agents make a common ID set by simulating a parallel Byzantine consensus algorithm and gather by using the common ID set.
In future work, it is interesting to consider the case that agents start at different times.
In the existing gathering algorithm \cite{Hirose2021}, when an agent starts the algorithm, it executes an exploring algorithm to wake up sleeping agents.
By this behavior, this algorithm creates an upper bound on the startup delay between good agents, and thus it deals with the startup delay in a similar way to simultaneous startup.
We will examine whether this approach can be taken for the proposed algorithm as well.
It is also worth studying a gathering algorithm using a Byzantine consensus algorithm with less than $9f+8$ agents.

\bibliographystyle{unsrt}
\bibliography{ref}

\end{document}